\documentclass{LMCS}

\usepackage{amsmath,amssymb}
\usepackage{DMC,STLC,Translation,DCCpc}
\usepackage{latexsym}
\usepackage[all]{xy}
\usepackage{hyperref}

\newcommand{\dc}{$\lambda^{[\,]}$}
\newcommand{\DC}{Sealing Calculus}
\newcommand{\DCCpc}{\ensuremath{\mathrm{DCC_{pc}}}}
\newcommand{\dco}{$\lambda^{[\,]}$}
\newcommand{\ctxeq}[4]{%
\ensuremath{#2 \stackrel{\mathrm{\scriptscriptstyle ctx}}=_{#1} #3 : #4}}
\newcommand{\nfeq}[2]{%
\ensuremath{#1 \stackrel{\mathrm{\scriptscriptstyle nf}}{=} #2}}
\newcommand{\defeq}{\stackrel{\mathrm{def}}{=}}
\newtheorem*{mycase}{Case}{\itshape}{\rmfamily}
\newtheorem*{mysubcase}{Subcase}{\itshape}{\rmfamily}

\newif\ifdraft\drafttrue
\newif\iffull\fulltrue

\def\doi{4 (3:10) 2008}
\lmcsheading%
{\doi}
{1--31}
{}
{}
{Sep.~25, 2007}
{Sep.~20, 2008}
{}   

\begin{document}
\title[Proving Noninterference %
       by a Fully Complete Translation]%
      {Proving Noninterference %
       by a Fully Complete Translation to %
       the Simply Typed $\lambda$-calculus\rsuper*}
\author[N.~Shikuma]{Naokata Shikuma}
\address{Graduate School of Informatics, Kyoto University, Kyoto
606-8501 Japan}
\email{\{naokata,igarashi\}@kuis.kyoto-u.ac.jp}

\author[A.~Igarashi]{Atsushi Igarashi}



\titlecomment{{\lsuper*}An earlier version of the paper has appeared in the
  Proceedings of the 11th Annual Asian Computing Science Conference
  (ASIAN'06), Springer-Verlag LNCS 4435, pp. 302--316, 2006.}

\begin{abstract}
Tse and Zdancewic have formalized the notion of noninterference for
Abadi et al.'s DCC in terms of logical relations and given a proof of
noninterference by reduction to parametricity of System F.
Unfortunately, their proof contains errors in a key lemma that their
translation from DCC to System F preserves the logical relations
defined for both calculi.  In fact, we have found a counterexample for
it.  In this article, instead of DCC, we prove noninterference for
\emph{sealing calculus}, a new variant of DCC, by reduction to the
basic lemma of a logical relation for the simply typed
\(\lambda\)-calculus, using a \emph{fully complete} translation to the
simply typed \(\lambda\)-calculus.  Full completeness plays an
important role in showing preservation of the two logical relations
through the translation.  Also, we investigate relationship among
sealing calculus, DCC, and an extension of DCC by Tse and Zdancewic
and show that the first and the last of the three are equivalent.

\end{abstract}

\maketitle

\section{Introduction}
\label{sec:Intro}
\paragraph{Background.}
Dependency analysis is a family of static program analyses to trace
dependencies between inputs and outputs of a given program.  For
example, information flow analysis~\cite{DenningDenning77CACM},
binding-time analysis~\cite{JonesGomardSestoft93PEbook}, and call
tracking~\cite{TangJouvelot95PEPM} are its instances.  One of the most
important correctness criteria of the dependency analysis is called
\emph{noninterference} \cite{goguenmeseguer82ieee}, which roughly
means that, for any pair of program inputs that are equivalent from
the viewpoint of an observer at some dependency level (e.g., 
security level, binding-time), the outputs are
also equivalent for the observer.  Various techniques for type-based
dependency analyses have been proposed, especially, in the context of
language-based security~\cite{SabelfeldMyers03IEEE}.

Abadi et al.\ proposed a unifying framework called \emph{dependency core
calculus} (DCC) \cite{abadi99core} for type-based dependency analyses
for higher-order functional languages, and gave it a denotational model
whose idea comes from \emph{parametricity}
\cite{DBLP:conf/ifip/Reynolds83,wadler89theorems} of System
F~\cite{REYNOLDS74,Girard72} through other information flow analyses
\cite{heintze-nevin98slam,mizunoS92sec-algo}. They showed
noninterference for several type systems of concrete dependency analyses
by embedding them into DCC.

Recently, Tse and Zdancewic
\cite{steve-translating-jfp-draft,steve-translating-acm-conf,steve-translating-techreport}
studied the relationship between DCC and System F. 
First, they formalized the noninterference property for
recursion-free DCC by using a syntactic \emph{logical relation}
\cite{Mitchell96}---a family of type-indexed relations, defined by
induction on types, over programs---as the equivalence relations for
inputs and outputs, thereby generalizing the notion of noninterference
to higher-order inputs and outputs.  Then, they gave a proof of
noninterference by reducing it to the parametricity theorem, which was
also formalized in terms of syntactic logical relations, of System
F.  Their technical development is summarized as follows:
\begin{enumerate}
 \item Define a translation $\mathcal{F}$ from DCC to System F;
 \item Prove, by induction on the structure of types, that the
   translation is both sound and complete---that is, it preserves the
   logical relations in the sense that
\[
         \mDMCterm_1 \approx_D \mDMCterm_2 : \mDMCtype \iff
\mathcal{F}(\mDMCterm_1) \approx_F \mathcal{F}(\mDMCterm_2) : \mathcal{F}(\mDMCtype)
\]
where $t$ is a DCC type,
and $\approx_D$ and $\approx_F$ represent the logical relations for
DCC and System F, respectively; and

\item Prove noninterference by reduction to the parametricity theorem
  of System F, using the sound and complete translation above.
\end{enumerate}

Unfortunately, in the second step, their proof
\cite{steve-translating-jfp-draft,steve-translating-acm-conf,steve-translating-techreport}
contains an error%
\footnote{The latest version~\cite{steve-translating-jfp-draft} was
  submitted and accepted for publication, but, due to this flaw, has
  not been published yet.  The authors are fixing the problem
  (personal communication with the authors).}, which we will briefly
explain here.  Note first that, for function types $t_1 \to t_2$, the
logical relations are defined by: $e_1 \approx_x e_2 : t_1 \to t_2$ if
and only if $e_1\,e_1' \approx_x e_2\,e_2' : t_2$ for any $e_1'
\approx_x e_2' : t_1$ ($x$ stands for either $D$ or $F$) and that the
type translation is homomorphic for function types, namely
\(\mathcal{F}(\mDMCtype_1 \to \mDMCtype_2) = \mathcal{F}(\mDMCtype_1)
\to \mathcal{F}(\mDMCtype_2)\).  Then, consider the case where
\(\mDMCtype\) is a function type \(t_1 \to t_2\).
To show the left-to-right direction, we must show that
$\mathcal{F}(\mDMCterm_1)\,\mSTLCterm_1 \approx_F
\mathcal{F}(\mDMCterm_2)\,\mSTLCterm_2 : \mathcal{F}(\mDMCtype_2)$ for
any $\mSTLCterm_1 \approx_F \mSTLCterm_2 : \mathcal{F}(\mDMCtype_1)$,
from the assumption \(\mDMCterm_1 \approx_D \mDMCterm_2 : \mDMCtype_1
\to \mDMCtype_2\), but we get stuck because there is no applicable
induction hypothesis.  If there existed a DCC term \(e\) such that
\(\mathcal{F}(e) = M\) for any System F term \(M\) of type
\(\mathcal{F}(\mDMCtype)\)---in this case, we say a translation is
\emph{full}~\cite{Hasegawa00JFP}---then \(\mSTLCterm_1\) and
\(\mSTLCterm_2\) would be of the forms \(\mathcal{F}(\mDMCterm_1')\)
and \(\mathcal{F}(\mDMCterm_2')\), making it possible to apply an
induction hypothesis, and the whole proof would go through.  Their
translation, however, turns out \emph{not} to be full; we have
actually found a counterexample for the preservation of the
equivalence from the failure of the fullness (see
Section~\ref{sec:DCC} for more details).  So, although interesting,
this indirect proof method fails at least for the combination of DCC
and System F.  Note that the noninterference property itself could be
proved directly by induction on DCC typing.

\paragraph{Our Contributions.}  
In this paper, we prove noninterference by Tse and Zdancewic's method in
a slightly different setting: In order to obtain a fully complete
translation, we change the source language to a richer one, what we call
\DC{} (\dc{}), and use a simpler target language, namely the simply
typed \(\lambda\)-calculus $\STLC$.  Then, the basic lemma for logical
relations of $\STLC$ is used in place of the parametricity theorem.

\dc{} is a simply typed $\lambda$-calculus with the notion of sealing
and a simplification of a security calculus which Tse and Zdancewic
proposed as an extension of DCC (we call it \DCCpc{} throughout this
paper) ~
\cite{steve-translating-jfp-draft,steve-translating-acm-conf,steve-translating-techreport}.
A \dc{} term $\DMCLlamExp{\ell}{\mDMCterm}$ stands for sealing
$\mDMCterm$ with a level $\ell$, which is a degree of confidentiality
of the sealed data. The sealed data can be extracted by unsealing
$\DMCLappExp{\mDMCterm}{\ell}$. For example, let $\mDMCtermNF$ a
sealed boolean value, then
$\DMCLappExp{(\DMCLlamExp{\ell}{\mDMCtermNF})}{\ell}$ is evaluated to
$\mDMCtermNF$.  We control unsealing operations by a type system so
that only users with relevant authority can unseal.  In the type
system, e.g., we assign a sealing type $\GuardType{\ell}{\bool}$ to
$\DMCLlamExp{\ell}{\mDMCtermNF}$ for any user, but,
$\DMCLappExp{(\DMCLlamExp{\ell}{\mDMCtermNF})}{\ell}$ has type $\bool$
only for authorized users.  To take such a notion of ``authorized
users'' into account, a type judgment is augmented with information
about authority.

Then, we define a translation of \dc{} to \STLC{}
in the same way as Tse--Zdancewic's translation of DCC
\cite{steve-translating-jfp-draft,steve-translating-acm-conf,steve-translating-techreport}:
we encode $\DMCLlamExp{\ell}{\mDMCtermNF}$ and its type
$\GuardType{\ell}{\bool}$ by $\lambda$-abstraction
$\STLCLamExp{k}{\alpha_\ell}{\mDMCtermNF}$ and function type
$\alpha_\ell \to \bool$, respectively, where $\alpha_\ell$ is a type
variable.  Intuitively, a term $K$ of type $\alpha_\ell$, if
exists, will be a key of unsealing, that is, we can apply
$\STLCLamExp{k}{\alpha_\ell}{\mDMCtermNF}$ to $K$ and get the sealed
value $\mDMCtermNF$.  The existence of such a typable term $K$ of
$\alpha_\ell$ in \STLC{} corresponds to a user's authority to
unseal with $\ell$ in \dc.
Our translation is full and, hence, there is no problem to prove
noninterference property of \dc{} under Tse--Zdancewic's scenario
described above.

Our main technical contributions can be summarized as follows:
\begin{itemize}
 \item Development of a sound and fully complete translation from \dc{} to
       $\STLC$;
 \item A proof of the noninterference theorem
       of \dc{} by reduction to the basic lemma of \STLC; and
 \item A proof of equivalence between \dc{} and \DCCpc.
\end{itemize}
As for DCC, noninterference can be proved directly by straightforward
induction in a manner quite similar to the basic lemma of \STLC.  So,
the main interest would not be in the noninterference property itself
but, rather, in how semantics of different calculi can be related with
each other by translation.  The existence of a fully complete
translation means that \dc{} provides syntax rich enough to express
every denotation in the model (that is, $\STLC$).  The translation is
also \emph{fully abstract}, as our logical relation for \dc{}
coincides with its contextual equivalence.  Also, comparing
Tse--Zdancewic's translation of DCC with ours, we have found and show
that, in spite of simplification, \dc{} is actually equivalent to
\DCCpc{} mentioned above.  This result indicates that both calculi are
really improvements over DCC.

This article is an extended version of our previous
paper~\cite{shikumaI06trans}.  In addition to giving detailed proofs, we
have extended the earlier version of \dc{} by introducing ordering on
levels, as DCC or \DCCpc, making it easier to compare \dc{} with them.

\paragraph{Structure of the Paper.}
The rest of the paper is organized as follows.
Section \ref{sec:DMC} introduces \dc{} with its syntax, type system,
reduction, and logical relations and then the statement of the
noninterference theorem.
In Sections~\ref{sec:STLC} and \ref{sec:Translation} 
we introduce $\STLC$ and define a translation
from \dc{} to $\STLC$ and its inverse.
In Section \ref{sec:Proof}, we complete our proof of noninterference
by reducing it to the basic lemma of logical relations for $\STLC$.
Section \ref{sec:DCC} explains why Tse and Zdancewic's translation
from DCC to System F is neither full nor sound, 
introduces their extension \DCCpc, which recovers fullness,
and shows that \dc{} and \DCCpc{} are equivalent.
Finally, Section \ref{sec:Conclusion} gives concluding remarks.


\section{\texorpdfstring{\DC}{DC}}
\label{sec:DMC}
In this section, we define \emph{\dc}, which is the simply typed
\(\lambda\)-calculus with sealing.  

First, we will introduce two kinds of levels: \emph{data levels} and
\emph{observer levels}.  Intuitively, a data level represents a degree of
confidentiality of data, while an observer level represents a capability
of an observer (e.g., a user or a process) to access data. The observer
can access only data whose data level $\ell$ is \emph{lower than} (i.e.,
inside of the range of) his or her observer level $\barell$. Moreover,
he or she can just obtain information depending on such data.

Then, we will define the terms, type systems, and reduction semantics of
\dc{} and show some basic properties. As mentioned in the previous
section, we write $\DMCLlamExp{\ell}{\mDMCterm}$ for sealing a \dc{} term
\(\mDMCterm\) with a data level \(\ell\).
The sealed value can be
extracted by unsealing \(\DMCLappExp{\mDMCterm}{\ell}\), whose result
must not be leaked to any observer whose observer level is not higher
than $\ell$. We control such dependency
by the type system. In this system, information on the data
level $\ell$ used for sealing is attached to types of sealing
$\GuardType{\ell}{\mDMCtype}$; furthermore, type judgments, written
\(\DMCJudgeOne{\mDMCterm}{\mDMCtype}\), are augmented by an observer level
\(\barell\), which is also called a \emph{protection context} elsewhere
\cite{steve-translating-acm-conf,steve-translating-techreport,steve-translating-jfp-draft},
as well as by a typing context $\Gamma$, which is a (finite) mapping from
variables to types.  This
judgment means that the value of \(\mDMCterm\) has type \(\mDMCtype\) as
usual and, moreover, can be leaked to (any observer at) an observer level
higher than \(\barell\).

Finally, we will formalize equivalences for \dc{} and give the formal
statement of noninterference.  The equivalences are indexed
by observer levels.  In the definition, any two values sealed at
the same data level will always be considered equal, or
indistinguishable, unless the observer level is higher than the data
level; and then the noninterference amounts to saying that, given inputs
equal at a given observer level, a typable program yields equal outputs
(at the same level).  So, in other words, an observer level reflects how
much power one has to distinguish the extensional behavior of programs
by investigating the contents of (sealed) values returned by the
programs.

\subsection{Syntax}
Let $(\mathcal{L}, {\sqsubseteq})$ be a poset where $\mathcal{L}$ is a
finite set of data levels, ranged over by $\ell$, and ${\sqsubseteq}$ is a
partial order over $\mathcal{L}$.  The metavariable \(\barell\) ranges
over observer levels, which are finite subsets of data levels.  We will
often omit the qualifications ``data'' and ``observer'' for levels unless
there is no confusion.
Observer levels are pre-ordered as follows: $\barell_1
\sqsubseteq \barell_2$ if and only if, for any $\ell_1 \in \barell_1$,
there exists $\ell_2 \in \barell_2$ such that $\ell_1 \sqsubseteq
\ell_2$. We also abbreviate $\{\ell\} \sqsubseteq \barell$ to $\ell
\sqsubseteq \barell$.  
\begin{rem}
  The notions of authorities and levels in the early version of this
  article~\cite{shikumaI06trans} correspond to those of data and
  observer levels here.  A main difference is that authorities were
  not given an order but data levels are partially ordered as in DCC.
  We have changed them to follow the standard terminology but also
  introduce an explicit distinction between two kinds of
  levels---those of data and those of observers.
\end{rem}
\begin{rem}
  We could unify data and observer levels and use a lattice, which is
  more standard in security calculi~
  \cite{abadi99core,heintze-nevin98slam}, to define \dc{}, just as in
  (precisely speaking, an earlier
  version~\cite{steve-translating-acm-conf,steve-translating-techreport}
  of) Tse and Zdancewic's extension of DCC.  Nevertheless, we adopt a
  poset for data levels and the pre-ordered set induced from it for
  observer levels, because it would be rather complicated (and also
  tedious) to translate such a variant into \STLC.  Note that the
  observer levels can be viewed as a lattice by identifying any two
  elements that are greater than each other.
\end{rem}

Then, the types of \dc{} are defined as follows.
\begin{defi}[Types]
The set of \emph{types}, ranged over by
$\mDMCtype,\, \mDMCtype',\, \mDMCtype_1,\, \mDMCtype_2,\,\dots$,
is defined as follows:
\[
 \mDMCtype ::=  \DMCunit \mid \mDMCtype\to\mDMCtype \mid
 \mDMCtype \times \mDMCtype \mid
 \mDMCtype + \mDMCtype \mid \GuardType{\ell}{\mDMCtype}
\]
We call \GuardType{\ell}{\mDMCtype} a {\em sealing type}.
\end{defi}
%
We define the terms of \dc{}  below.  The metavariables $x$, $y$, and
$z$ (possibly with subscripts) range over the denumerable set of
\emph{variables}.
\begin{defi}[Terms]
The set of \emph{terms}, ranged over by
$\mDMCterm,\, \mDMCterm',\, \mDMCterm_1,\, \mDMCterm_2,\,\dots$,
is defined as follows:
 \begin{align*}
  \mDMCterm &::= x \mid ()
        \mid \DMClamExp{x}{\mDMCtype}{\mDMCterm}
        \mid \mDMCterm\, \mDMCterm
        \mid \DMCpair{\mDMCterm}{\mDMCterm}
        \mid \DMCpiOne{\mDMCterm} \mid \DMCpiTwo{\mDMCterm}
        \mid \DMCiotaOne{\mDMCterm} \mid \DMCiotaTwo{\mDMCterm} \\
    & \mid \DMCcasExp{\mDMCterm}{\mDMCterm}{\mDMCterm}
      \mid \DMCLlamExp{\ell}{\mDMCterm}
      \mid \DMCLappExp{\mDMCterm}{\ell}
 \end{align*}
\end{defi}
Terms of \dc{} include variable, the unit value,
\(\lambda\)-abstraction, application, pairing, projection, injection,
and case analysis.
As usual, \(x\) is bound in \(\mDMCterm\) of
\(\DMClamExp{x}{\mDMCtype}{\mDMCterm}\) and \(x_1\) and \(x_2\) are
bound in \(e_1\) and \(e_2\) of
\DMCcasExp{\mDMCterm_0}{\mDMCterm_1}{\mDMCterm_2}, respectively.  We
say, for $\DMCLlamExp{\ell}{\mDMCterm}$, $\mDMCterm$ is {\em sealed at}
$\ell$, and call $\DMCLlamExp{\ell}{\mDMCterm}$ and
$\DMCLappExp{\mDMCterm}{\ell}$ a \emph{sealing term} and
an \emph{unsealing term}, respectively.
In this paper, \(\alpha\)-conversions are defined in a customary
manner and implicit \(\alpha\)-conversions are assumed to make all the
bound variables distinct from other (bound and free) variables.

\subsection{Type System}
As mentioned above, the form of type judgment of \dc{} is
\DMCJudgeOne{\mDMCterm}{\mDMCtype}.  This judgment is read as
``$\mDMCterm$ is given type $\mDMCtype$ at observer level $\barell$
under context $\DMCContext$.''  The intuition is that the computation of
\(e\) depends on only data levels lower than \(\barell\), and so the
information on its value can be leaked only to an observer level
\(\barell'\), which is higher than \(\barell\).

The typing rules of \dc{} are given as follows:

\DMCTypingRulesOneColumn
All the rules but the last two are straightforward.  The rule (\RDMCLabs) for
sealing means that, by sealing with \(\ell\), 
it is legal to leak \(\DMCLlamExp{\ell}{\mDMCterm}\) to an observer level
which is not higher than \(\ell\): at such an observer level, however,
\(\mDMCterm\) cannot be unsealed, as is shown in the rule (\RDMCLapp)
for unsealing.
\begin{exa}
The following judgment
\begin{align*}
 \cdot\,;\,\barell\,\seqsym\,& \DMClamExp{x}{\GuardType{\ell_1}{\mDMCtype_1 + \mDMCtype_2}}
   {\DMCLlamExp{\ell_2}
    {\DMCcasExp{\DMCLappExp{x}{\ell_1}}
     {\DMCiotaOne{\DMCLlamExp{\ell_3}{x_1}}}
     {\DMCiotaTwo{\DMCLlamExp{\ell_3}{x_2}}}}} \\
 & :\,\GuardType{\ell_1}{\mDMCtype_1 + \mDMCtype_2}
  \to \GuardType{\ell_2}{\GuardType{\ell_3}{\mDMCtype_1} + \GuardType{\ell_3}{\mDMCtype_2}}
\end{align*}
is derivable if and only if $\ell_1 \sqsubseteq \barell \cup
\{\ell_2\}$, which is required at unsealing $x$ of
$\GuardType{\ell_1}{\mDMCtype_1 + \mDMCtype_2}$ with $\ell_1$---the
observer level there is $\barell \cup \{\ell_2\}$ and must be higher
than the data level $\ell_1$.
\end{exa}

The type constructor \(\GuardType{\ell}{\cdot}\) is
very similar to the (indexed) monadic type constructor
\(T_\ell\) in DCC~\cite{abadi99core}.  In fact, the logical relations we
will define for \dc{} are essentially the same as those defined for DCC
and a main idea of the translation from \dc{} to \STLC{} is also the
same as that from DCC to System
F~\cite{steve-translating-jfp-draft,steve-translating-acm-conf,steve-translating-techreport}.
Nevertheless, we have chosen a different symbol as the monadic bind
construct is no longer used in \dc{} and, as a result, the type system
is fairly different from DCC.  We will give a more detailed comparison
with DCC (and its
extension~\cite{steve-translating-jfp-draft,steve-translating-acm-conf,steve-translating-techreport})
in Section~\ref{sec:DCC}.
\subsection{Reduction}
The \emph{reduction relation} for \dc{} is written $\mDMCterm \longrightarrow
\mDMCterm'$, which expresses that 
$\mDMCterm$ is \emph{reduced} to $\mDMCterm'$ by applying one of the
following rules to a subterm of $\mDMCterm$.
\[
\begin{array}{rcl}
  (\DMClamExp{x}{\mDMCtype}{\mDMCterm_1})\, \mDMCterm_2 & 
  \longrightarrow &
  [\mDMCterm_2 / x]\mDMCterm_1 \\
  \DMCpii{\DMCpair{\mDMCterm_1}{\mDMCterm_2}} &
  \longrightarrow & \mDMCterm_i \\
  \DMCcasExp{\DMCiotai{\mDMCterm}}{\mDMCterm_1}{\mDMCterm_2} & 
  \longrightarrow & [\mDMCterm / x_i]\mDMCterm_i \\
   \DMCLappExp{(\DMCLlamExp{\ell}{\mDMCterm})}{\ell} &
   \longrightarrow & \mDMCterm \\
\end{array}
\]
We write $[\mDMCterm / x]$ for a capture-avoiding substitution of
$\mDMCterm$ for the free occurrences of variable $x$.  All rules are
straightforward.  The last rule says that the term sealed by \(\ell\)
is opened by the same level.  In what follows, we use \(\mDMCtermNF\)
for \emph{normal forms}, that is, terms which cannot be reduced
anymore.  Note that
$\DMClamExp{x}{\mDMCtype}{\DMCLappExp{(\DMCLlamExp{\ell}{x})}{\ell}}$
is \emph{not} a normal form, since the reduction is full, that is,
even a redex under \(\lambda\)-abstraction can be reduced.  We write
$\longrightarrow^*$ for the reflexive transitive closure of
$\longrightarrow$.
\subsection{Basic Properties}

We list some basic properties of \dc.  The first lemma below means that,
if \(\mDMCterm\) is well typed at some observer level, then it is
also well typed at a higher level.

\begin{lem}[Observer Level Monotonicity]
\label{lem:LM}
If \DMCJudge{\DMCContext}{\barell_1}{\mDMCterm}{\mDMCtype} and
 $\barell_1 \sqsubseteq \barell_2$, 
then \DMCJudge{\DMCContext}{\barell_2}{\mDMCterm}{\mDMCtype},
and the derivations of these judgments have the same size.
\end{lem}
\proof
By induction on the derivation of
 \DMCJudge{\DMCContext}{\barell_1}{\mDMCterm}{\mDMCtype,} using the fact
 that $\barell_1 \cup \barell \sqsubseteq \barell_2 \cup \barell$ if
 $\barell_1 \sqsubseteq \barell_2$.
\qed

\begin{lem}[Substitution Property]
 \label{lem:SP}
 If \DMCJudgeOne{\mDMCterm}{\mDMCtype} and
 \DMCJudge{\Gamma,\, \DMCDec{x}{\mDMCtype}}{\barell}{\mDMCterm'}{\mDMCtype'},
 then \DMCJudgeOne{[\mDMCterm/x]\mDMCterm'}{\mDMCtype'}
\end{lem}
\proof
 By induction on the derivation of
 \DMCJudge{\Gamma,\, \DMCDec{x}{\mDMCtype}}{\barell}{\mDMCterm'}{\mDMCtype'},
 using Lemma \ref{lem:LM}.
\qed

The following three theorems are standard.
\begin{thm}[Subject Reduction]
 If \DMCJudgeOne{\mDMCterm}{\mDMCtype}
 and $\mDMCterm\, \longrightarrow\, \mDMCterm'$,
 then \DMCJudgeOne{\mDMCterm'}{\mDMCtype}.
\end{thm}
\proof
 By induction on the derivation of
 \DMCJudgeOne{\mDMCterm}{\mDMCtype},
 using Lemmas \ref{lem:LM} and \ref{lem:SP}.
\qed

\begin{thm}[Strong Normalization]
\label{theo:SN}
 If \DMCJudgeOne{\mDMCterm}{\mDMCtype}, 
 then $\mDMCterm$ is \emph{strongly normalizing}, that is, there is no infinite
 sequence of reductions which starts from $\mDMCterm$.
\end{thm}
\iffull
\proof
 Define a translation from \dc{} into the simply typed $\lambda$-calculus
 as follows:
 \begin{align*}
  (\GuardType{\ell}{\mDMCtype})^\dag &= \DMCunit \to \mDMCtype^\dag \\
  (\DMCLlamExp{\ell}{\mDMCterm})^* &=
    \DMClamExp{\_}{\DMCunit}{\mDMCterm^*} \\
  (\DMCLappExp{\mDMCterm}{\ell})^* &= \mDMCterm^* \, ().
 \end{align*}
 This translation preserves typing and maps a reduction 
 $\mDMCterm_1 \longrightarrow \mDMCterm_2$
 to $\mDMCterm_1^* \longrightarrow^+ \mDMCterm_2^*$, where
 $\longrightarrow^+$ is the transitive closure of $\longrightarrow$.
 So, from strong normalization for the simply typed $\lambda$-calculus
 (see, e.g., \cite{Mitchell96}), we conclude
 one for \dc.
\qed
\fi
\begin{thm}[Church-Rosser Property]
 If \DMCJudgeOne{\mDMCterm}{\mDMCtype} and 
 $\mDMCterm\, \longrightarrow^*\, \mDMCterm_1$
 and $\mDMCterm\, \longrightarrow^*\, \mDMCterm_2$,
 then there exists a term $\mDMCterm'$ 
 such that $\mDMCterm_i\, \longrightarrow^*\, \mDMCterm'$
 {\rm(}$i = 1, 2${\rm)}.
\end{thm}
\proof By Theorem \ref{theo:SN} and Newman's Lemma \cite{newman42amath},
it suffices to show that the reduction is weakly confluent: If
\DMCJudgeOne{\mDMCterm}{\mDMCtype} and $\mDMCterm\, \longrightarrow\,
\mDMCterm_1$ and $\mDMCterm\, \longrightarrow\, \mDMCterm_2$, then there
exists a term $\mDMCterm'$ such that $\mDMCterm_i\, \longrightarrow^*\,
\mDMCterm'$ {\rm(}$i = 1, 2${\rm)}.  This is easy.  
\qed
\subsection{Contextual Equivalence, Noninterference, and Logical Relations}
Now we formalize equivalence of terms from the viewpoint of an observer
at a given level as \emph{contextual equivalence}, and then state a
formalization of noninterference.

We say that $\mDMCterm_1$ and $\mDMCterm_2$ are contextually equivalent at
observer level \(\barell\) if $C[\mDMCterm_1]$ and $C[\mDMCterm_2]$ are
evaluated to the same value for any context $C[\cdot]$ typed at
\(\barell\).  Note that the equivalence is indexed by an observer level.
We define contextual equivalence
$\stackrel{\textrm{\tiny ctx}}=_{\barell}$
as follows:
\begin{defi}[Contextual Equivalence for \dc]
  Assume that \DMCJudge{\cdot}{\barell}{\mDMCterm_i}{\mDMCtype} for $i =
 1, 2$ (we write \(\cdot\) for the empty variable context).  The
 relation \ctxeq{\barell}{\mDMCterm_1}{\mDMCterm_2}{\mDMCtype} is
 defined by: \ctxeq{\barell}{\mDMCterm_1}{\mDMCterm_2}{\mDMCtype} if and
 only if \nfeq{f \mDMCterm_1}{f \mDMCterm_2} for any $f$ such that
 \DMCJudge{\cdot}{\barell} {f}{\mDMCtype \to \bool.}  Here,
 \nfeq{\mDMCterm}{\mDMCterm'} means that $\mDMCterm$ and $\mDMCterm'$
 have the same normal form and \bool{} stands for $\DMCunit + \DMCunit$.
\end{defi}
Here we use functions as contexts without loss of generality, because,
by Strong Normalization and Church-Rosser, $C[\mDMCterm]$ and
$(\DMClamExp{x}{\mDMCtype}{C[x]})\, \mDMCterm$ has a unique normal form,
where $\mDMCtype$ is the type of $\mDMCterm$.

The following proposition shows that an observer level in the
contextual equivalence reflects an observer's distinguishability, in
other words, that an observer at a lower level can distinguish no more
terms than another at a higher.
\begin{prop}[]
 Assume that \DMCJudge{\cdot}{\barell_1}{\mDMCterm_i}{\mDMCtype} for $i =
  1, 2$.  If $\pi_1 \sqsubseteq \pi_2$ and
  \ctxeq{\barell_2}{\mDMCterm_1}{\mDMCterm_2}{\mDMCtype}, then
  \ctxeq{\barell_1}{\mDMCterm_1}{\mDMCterm_2}{\mDMCtype}.
\end{prop}
\proof
Take a function $f$ such that \DMCJudge{\cdot}{\barell_1}%
{f}{\mDMCtype \to \bool.} By Observer Level Monotonicity (Proposition
\ref{lem:LM}), \DMCJudge{\cdot}{\barell_2}{f}{\mDMCtype \to \bool}
and \DMCJudge{\cdot}{\barell_2}{\mDMCterm_i}{\mDMCtype} $(i = 1, 2)$.
By assumption, \nfeq{f \mDMCterm_1}{f \mDMCterm_2}, and so
\ctxeq{\barell_1}{\mDMCterm_1}{\mDMCterm_2}{\mDMCtype}.
\qed

We use \(\gamma\) to represent a simultaneous substitution of terms for
variables and write \ctxeq{\barell}{\gamma_1}{\gamma_2}{\DMCContext} if
\(\dom(\gamma_1) = \dom(\gamma_2) = \dom(\DMCContext)\) and
\ctxeq{\barell}{\gamma_1(x)}{\gamma_2(x)}{\DMCContext(x)} for any \(x\in
\dom(\gamma_1)\).  Then, the noninterference is defined as
follows:
\begin{defi}[Noninterference]
Take $\mDMCterm$ such that \DMCJudgeOne{\mDMCterm}{\mDMCtype}.
\emph{The well typed term $\mDMCterm$ satisfies noninterference}, if and only
if, \ctxeq{\barell}{\gamma_1(\mDMCterm)}{\gamma_2(\mDMCterm)}
{\mDMCtype} for any $\gamma_1$ and $\gamma_2$ such that
\ctxeq{\barell}{\gamma_1}{\gamma_2}{\DMCContext}.
\end{defi}
As mentioned before, noninterference means that, for any
pair of program inputs that are equivalent from the viewpoint of an
observer at some security level, the outputs are also equivalent for
the observer.  Here, substitutions \(\gamma_1\) and \(\gamma_2\) play
roles of equivalent inputs to program \(e\).  So, this property
specifies the correctness of the type system as a dependency analysis.

Though we want to show that any well typed term satisfies the
noninterference above, this is hard due to the following generally-known
fact: it is difficult, in general, to show given two terms are
contextually equivalent.  The reason is that we must take account of
\emph{all} contexts but proof by induction on the structure of contexts
does not usually work.

To solve this problem, we use the well-known technique of \emph{logical
relations}~\cite{Mitchell96,Plotkin80Curry}, which will be shown to be
equivalent to the contextual equivalences, and state the noninterference
theorem in terms of the logical relations.

As the contextual equivalence above, the logical relations (for close
terms and closed normal forms) are indexed by observer levels as well
as types. A judgment
\DMCLogicalRelation{\barell}{\mDMCterm_1}{\mDMCterm_2}{\mDMCtype}
means that closed terms $\mDMCterm_1$ and $\mDMCterm_2$ of type
$\mDMCtype$ are logically related at observer level $\barell$.
Similarly,
\DMCLogicalRelationNF{\barell}{\mDMCtermNF_1}{\mDMCtermNF_2}{\mDMCtype}
means that closed normal forms $\mDMCtermNF_1$ and $\mDMCtermNF_2$ of
$\mDMCtype$ are logically related at $\barell$.  We assume
\(\DMCJudge{\cdot}{\barell}{\mDMCterm_i}{t}\) and
\(\DMCJudge{\cdot}{\barell}{\mDMCtermNF_i}{t}\) for \(i = 1, 2\).

\begin{defi}[Logical Relations for \dc]
\label{def:LRDC}
The relations
 \DMCLogicalRelationNF{\barell}{\mDMCtermNF_1}{\mDMCtermNF_2}{\mDMCtype}
 and
 \DMCLogicalRelation{\barell}{\mDMCterm_1}{\mDMCterm_2}{\mDMCtype}
 are defined by the following rules:

\DMCLogicalRelationRules
\end{defi}

Most rules are straightforward.  In the rule (\RDMCLRFun), the premise
is the abbreviation of the following: $\forall e_1.~ \forall e_2.~
\DMCLogicalRelation{\barell}{e_1}{e_2}{t_1} \Rightarrow
\DMCLogicalRelation{\barell}{v_1 e_1}{v_2 e_2}{t_2}$.  There are two
rules for \DMCLogicalRelationNF{\barell}%
{\DMCLlamExp{\ell}{\mDMCtermNF_1}}%
{\DMCLlamExp{\ell}{\mDMCtermNF_2}}%
{\GuardType{\ell}{\mDMCtype}}.  When $\ell \sqsubseteq \barell$, an
observer at $\barell$ can examine $\mDMCtermNF_i$ by unsealing
\DMCLlamExp{\ell}{\mDMCtermNF_i} $(i = 1, 2)$, so these sealing terms
are equivalent only when its contents are equivalent.  Otherwise, the
observer cannot distinguish them at all and those terms are always
regarded equivalent.

\begin{exa}
\label{ex:LEone} We write $\mathtt{true}$ and $\mathtt{false}$,
respectively, for $\DMCiotaOne{()}$ and $\DMCiotaTwo{()}$.  Let
$\levL$ and $\levH$ data levels
and suppose that $\levL$ is
strictly lower than $\levH$.  Take any $\mDMCterm_i$ such that
$\DMCJudge{\cdot}{\levL}{\mDMCterm_i}{\GuardType{\levH}{\bool}}$
$(i = 1, 2)$. Then
\DMCLogicalRelation{\levL}{\mDMCterm_1}{\mDMCterm_2}
{\GuardType{\levH}{\bool}}. This follows from the facts that
\DMCLogicalRelationNF{\levL}{\DMCLlamExp{\levH}{c_1}}
{\DMCLlamExp{\levH}{c_2}} {\GuardType{\levH}{\bool}} where
$c_1, c_2 \in \{\mathtt{true}, \mathtt{false}\}$ and that each
$\mDMCterm_i$ has either normal form
$\DMCLlamExp{\levH}{\mathtt{true}}$ or
$\DMCLlamExp{\levH}{\mathtt{false}}$. 
\end{exa}

We define
\DMCLogicalRelation{\barell}{\gamma_1}{\gamma_2}{\DMCContext} 
similarly to \ctxeq{\barell}{\gamma_1}{\gamma_2}{\DMCContext.}
Then, the noninterference theorem is stated as follows:
\begin{thm}[Noninterference]
\label{theo:NonInterference}
If \DMCJudgeOne{\mDMCterm}{\mDMCtype} and
\DMCLogicalRelation{\barell}{\gamma_1}{\gamma_2}{\DMCContext}, then
\DMCLogicalRelation{\barell}{\gamma_1(\mDMCterm)}{\gamma_2(\mDMCterm)}%
  {\mDMCtype}.
\end{thm}
We will give a proof in Section \ref{sec:Proof}.

\begin{exa}
 Here, we use the same notations as Example \ref{ex:LEone}. Take a
 function $f$ such that \DMCJudge{\cdot}{\levL}{f}
 {\GuardType{\levH}{\bool} \to \GuardType{\levL}{\bool}}.  Now we will
 show that $f$ is a constant function. By the theorem above,
 \DMCLogicalRelation{\levL}{f}{f}{\GuardType{\levH}{\bool} \to
 \GuardType{\levL}{\bool}}.  From (\RDMCLRTerm), the discussion in Example
 \ref{ex:LEone} and (\RDMCLRFun), \DMCLogicalRelation{\levL}{f e_1}{f
 e_2} {\GuardType{\levL}{\bool}}. $f \mDMCterm_i$ has a normal form
 $\DMCLlamExp{\levL}{c_i}$ where some $c_i \in \{\mathtt{true},
 \mathtt{false}\}$ $(i = 1, 2)$ and, by (\RDMCLRTerm),
 \DMCLogicalRelationNF{\levL}{\DMCLlamExp{\levL}{c_1}}
 {\DMCLlamExp{\levL}{c_2}}{\GuardType{\levL}{\bool}}. So, by
 (\RDMCLRLabelTwo), $c_1 = c_2$, which means that $f$ always returns a
 constant value.
\end{exa}

Also, from the noninterference theorem (Theorem
\ref{theo:NonInterference}), it follows that
the logical relations exactly coincide with the contextual equivalences
above, and hence, in terms of the latter as well as the former,
the noninterference theorem also holds.
\begin{thm}[]
\label{theo:coin}
\DMCLogicalRelation{\barell}{\mDMCterm_1}{\mDMCterm_2}{\mDMCtype}
if and only if
\ctxeq{\barell}{\mDMCterm_1}{\mDMCterm_2}{\mDMCtype.}
\end{thm}
\proof
First, we show the right from the left. Suppose that
\DMCLogicalRelation{\barell}{\mDMCterm_1}{\mDMCterm_2}{\mDMCtype.}
Take arbitrary $f$ such that \DMCJudge{\cdot}{\barell}%
{f}{\mDMCtype \to \bool.} By Noninterference Theorem,
\DMCLogicalRelation{\barell}{f}{f}{\mDMCtype \to \bool,} and by
(\RDMCLRTerm) and (\RDMCLRFun), \DMCLogicalRelation{\barell}{f
  \mDMCterm_1}%
{f \mDMCterm_2}{\bool}. By (\RDMCLRTerm), (\RDMCLRInj) and
(\RDMCLRUnit), \nfeq{f \mDMCterm_1}{f \mDMCterm_2} and hence
\ctxeq{\barell}{\mDMCterm_1}{\mDMCterm_2}{\mDMCtype.}

Next, we prove the converse above by induction on the structure of
$\mDMCtype$. Assume that \ctxeq{\barell}{\mDMCterm_1}{\mDMCterm_2}{\mDMCtype.}
We show only the main cases:
\begin{mycase}[$\mDMCtype = \mDMCtype_1 \to \mDMCtype_2$]
Take arbitrary $\mDMCterm_1'$ and $\mDMCterm_2'$ such that
\DMCLogicalRelation{\barell}{\mDMCterm_1'}{\mDMCterm_2'}{\mDMCtype_1.}
By the left-to-right of Theorem \ref{theo:coin}
(which has been already shown in the first part of this proof),
\ctxeq{\barell}{\mDMCterm_1'}{\mDMCterm_2'}%
{\mDMCtype_1.} Take arbitrary $f$ such that \DMCJudge{\cdot}{\barell}%
{f}{\mDMCtype_2 \to \bool,} then \nfeq{f (\mDMCterm_1 \mDMCterm_1')}%
{f (\mDMCterm_1 \mDMCterm_2')} because \ctxeq{\barell}{\mDMCterm_1'}
{\mDMCterm_2'}{\mDMCtype_1.}
Also, by assumption, 
\nfeq{f (\mDMCterm_1 \mDMCterm_2')}{f (\mDMCterm_2 \mDMCterm_2'),}
and hence \nfeq{f (\mDMCterm_1 \mDMCterm_1')}{f (\mDMCterm_2
 \mDMCterm_2')} by transitivity of $=_\mathrm{nf}$.
So, \ctxeq{\barell}{\mDMCterm_1 \mDMCterm_1'}%
{\mDMCterm_2 \mDMCterm_2'}{\mDMCtype_2,} and by the induction
hypothesis for $\mDMCtype_2$, \DMCLogicalRelation{\barell}%
{\mDMCterm_1 \mDMCterm_1'}{\mDMCterm_2 \mDMCterm_2'}{\mDMCtype_2,}
therefore \DMCLogicalRelation{\barell}{\mDMCterm_1}{\mDMCterm_2}%
{\mDMCtype_1 \to \mDMCtype_2.}
\end{mycase}
\begin{mycase}[$\mDMCtype = \GuardType{\ell}{\mDMCtype_1}$]
We have two subcases according to whether $\ell \sqsubseteq \barell$ or not.
If $\ell \sqsubseteq \barell$, then, by Strong Normalization (Theorem
\ref{theo:SN}), there are normal forms $\mDMCtermNF_1$ and
 $\mDMCtermNF_2$ such that \DMCJudge{\cdot}{\barell}%
{\mDMCtermNF_i}{\mDMCtype_1} and $\mDMCterm_i \longrightarrow^*
 \DMCLlamExp{\ell}{\mDMCtermNF_i}$ for $i = 1, 2$.
Then, it must be the case that
\ctxeq{\barell}{\mDMCtermNF_1}{\mDMCtermNF_2}{\mDMCtype_1.}
(Otherwise, there would be a term $f$ such that
 \DMCJudge{\cdot}{\barell}{f}{\mDMCtype_1 \to \bool}
and $f \mDMCtermNF_1 \neq_\mathrm{nf} f \mDMCtermNF_2$. 
Let $f'$ be $\DMClamExp{x}{\GuardType{\ell}{\mDMCtype_1}}%
{f \DMCLappExp{x}{\ell}}$, then \DMCJudge{\cdot}{\barell}{f'}%
{\GuardType{\ell}{\mDMCtype_1} \to \bool} and $f' \mDMCterm_1
 \neq_\mathrm{nf} f' \mDMCterm_2$, and hence,
 $\mDMCterm_1 \neq_{\mathrm{ctx}}^{\barell} \mDMCterm_2 : 
\GuardType{\ell}{\mDMCtype_1}$, but this is a contradiction.)
Applying the induction hypothesis for $\mDMCtype_1$, 
\DMCLogicalRelation{\barell}{\mDMCtermNF_1}{\mDMCtermNF_2}{\mDMCtype_1,}
which is equivalent to \DMCLogicalRelationNF{\barell}{\mDMCtermNF_1}%
{\mDMCtermNF_2}{\mDMCtype_1,} so \DMCLogicalRelation{\barell}{\mDMCterm_1}%
{\mDMCterm_2}{\GuardType{\ell}{\mDMCtype_1}.} The case $\ell \not\sqsubseteq \barell$ is
trivial.
\qed
\end{mycase}


\section{The Simply Typed \texorpdfstring{$\lambda$}{lambda}-calculus}
\label{sec:STLC}
We review  the simply typed \(\lambda\)-calculus \(\STLC\)
briefly with logical relations for it.

\subsection{Definition of \texorpdfstring{\STLC}{STLC}}
\STLC{} introduced here is a standard one with unit, base, function,
product, and sum types.  We assume that base types, written
\(\alpha_\ell\) (\(\ell \in \mathcal{L}\)), have one-to-one
correspondence with data levels.  We use metavariables \(M\) for terms
and \(A\) for types.  The syntax of \STLC{} is given as follows:
\begin{center}
\STLCSyntax
\end{center}
Note that base type $\alpha_\ell$ has neither constants nor closed
terms.  The reason is that, as mentioned in Section \ref{sec:Intro}, we
will use a term of type $\alpha_\ell$ as a key for opening a sealing at
level $\ell$ and such a key should be permitted only to privileged
users. See Section \ref{sec:Translation} for details.

The form of type judgment of $\STLC$ is
$\STLCJudgeOne{\mSTLCterm}{\mSTLCtype}$, where $\STLCContext$ is a
(finite) mapping from variables to $\STLC$ types.  The typing rules are
given as follows:
\STLCTypingRules

The reduction of \STLC{} terms consists of standard \(\beta\)-reduction
\[
\begin{array}{rcl}
  (\STLCLamExp{x}{\mSTLCtype}{\mSTLCterm_1})\, \mSTLCterm_2 & 
  \longrightarrow &
  [\mSTLCterm_2 / x]\mSTLCterm_1 \\
  \STLCPii{\STLCpair{\mSTLCterm_1}{\mSTLCterm_2}} &
  \longrightarrow & \mSTLCterm_i \\
  \STLCCasExp{\STLCiotai{\mSTLCterm}}{\mSTLCterm_1}{\mSTLCterm_2} & 
  \longrightarrow & [\mSTLCterm / x_i]\mSTLCterm_i \\
\end{array}
\]
 and the following commutative conversion.
{\typicallabel{}
\CommutativeConversions
}
As in \dc, the reduction for \STLC{} is full, too.
Here, we write \(\FV(\mSTLCterm)\) for the set of free variables in
\(\mSTLCterm\).  In what follows, we use \(\mSTLCtermNF\) for normal
forms. For example, by the first and second commutative conversion rules,
\begin{align*}
& \STLCLamExp{z}{\STLCunit + \STLCunit}{\STLCPii{\STLCCasExp{z}{y_1}{y_2} z}} \\
& \longrightarrow \STLCLamExp{z}{\STLCunit + \STLCunit}{\STLCPii{\STLCCasExp{z}{y_1 z}{y_2 z}}} \\
& \longrightarrow \STLCLamExp{z}{\STLCunit + \STLCunit}{\STLCCasExp{z}{\STLCPii{y_1 z}}{\STLCPii{y_2 z}}},
\end{align*}
which is a normal form.

The resulting calculus (with commutative conversion) satisfies the
standard properties of subject reduction, Church-Rosser, and strong
normalization~\cite{deGroote02IC}.  We say (the type derivation
\STLCJudgeOne{\mSTLCterm}{\mSTLCtype} of) a term satisfies the
subformula property when any type in the derivation is a subexpression
of either \(\mSTLCtype\) or a type occurring in \(\STLCContext\).  Then, any
well typed term can reduce to the one that satisfies the subformula
property as in the theorem below, which makes it easy to ensure the
fullness of the translation.

\begin{thm}[Subformula Property]
\label{theo:SubForm}
If \(\STLCJudgeOne{\mSTLCterm}{\mSTLCtype}\), then there exists a normal
form $\mSTLCtermNF$ such that \(\mSTLCterm \longrightarrow^*
\mSTLCtermNF\) and \(\STLCJudgeOne{\mSTLCtermNF}{\mSTLCtype}\),
which satisfies the subformula property.
Also, all the subderivations satisfy the subformula property.
\end{thm}
\begin{rem}
Commutative conversion is necessary for the above theorem to hold.
Without commutative conversion,
\[
  \STLCLamExp{x}{\STLCunit + \STLCunit}%
                    {(\STLCCasExp{x}{\STLCLamExp{y}{\STLCunit}{()}}%
                       {\STLCLamExp{y}{\STLCunit}{()}})
                     \,()}%
\]
of type \(\STLCunit + \STLCunit \to \STLCunit\) would be a normal form,
which does not satisfy the subformula property, because a subterm
\(\STLCLamExp{y}{\STLCunit}{()}\) has type \(\STLCunit \to
\STLCunit\), which does not occur in \(\STLCunit + \STLCunit \to
\STLCunit\).  This theorem also requires full reduction, which allows
any redex (even under \(\lambda\)) to reduce.
\end{rem}
As mentioned above, we will view terms of type $\alpha_\ell$ as keys.
What really matters in the development below is \emph{whether} any key
of a given type exists or not and it is is not significant what kind
of keys exist.  Thus we identify all keys by introducing a (typed)
equivalence relation $\AxiomaticEquiv{\STLCContext}%
{\mSTLCterm_1}{\mSTLCterm_2}{\mSTLCtype}$.
\begin{defi}
The relation  $\AxiomaticEquiv{\STLCContext}%
   {\mSTLCterm_1}{\mSTLCterm_2}{\mSTLCtype}$
is defined as the least relation closed under the rules below:

\AxiomaticEquivRules
\end{defi}
\noindent{}
The rule (\AERKey) signifies that all keys are identified.
Clearly, \AxiomaticEquiv{\STLCContext}{\mSTLCterm}{\mSTLCterm}{\mSTLCtype} is
equivalent to \STLCJudgeOne{\mSTLCterm}{\mSTLCtype}.
\begin{lem}[$\equiv$ is Equivalence]
\label{lem:axeq}
Given \(\STLCContext\) and \(\mSTLCtype\), the binary relation
 $\AxiomaticEquiv{\STLCContext}{\cdot}{\cdot}{\mSTLCtype}$ on terms
 is an equivalence relation, that is, reflexive, symmetric, and transitive.
\end{lem}
\proof
Easy.
\qed
The following lemma says that two terms which differ only in subterms
of type $\alpha_\ell$ are equivalent via $\equiv$.
\begin{lem}
\label{lem:appxAE}
 Assume that \STLCJudgeOne{\mSTLCterm}{\mSTLCtype}.
 Take an occurrence $\mSTLCterm_1$ of type $\alpha_\ell$ in
 $\mSTLCterm$.
 Suppose that $\mSTLCterm_1$
 \emph{freely occurs in $\mSTLCterm$}, that is, no free variable of
 $\mSTLCterm_1$  is bound in the occurrence.
 If \STLCJudgeOne{\mSTLCterm_2}{\alpha_{\ell}}, then
 \AxiomaticEquiv{\STLCContext}{\mSTLCterm}%
{[\mSTLCterm_2/\mSTLCterm_1]\mSTLCterm} {\mSTLCtype,} where
$[\mSTLCterm_2/\mSTLCterm_1]\mSTLCterm$ is a result of
capture avoiding replacement of the occurrence $\mSTLCterm_1$ in $\mSTLCterm$
by $\mSTLCterm_2$.
In general, this holds for simultaneous replacing too.
\end{lem}
\proof
By induction on the derivation of \STLCJudge{\STLCContext}{\mSTLCterm}{\mSTLCtype.}
\qed

\subsection{Logical Relations for \texorpdfstring{\STLC}{STLC}}

We define syntactic logical relations for $\STLC$ in the standard
manner.  As for \dc{}, there are relations for (this time, possibly
open) terms and normal forms, written
\STLCLogicalRelation{\STLCContext}%
{\mSTLCterm_1}{\mSTLCterm_2}{\mSTLCtype} (read ``terms $\mSTLCterm_1$
and $\mSTLCterm_2$ of type $\mSTLCtype$ are logically related under
context $\STLCContext$'') and \STLCLogicalRelationNF{\STLCContext}%
{\mSTLCtermNF_1}{\mSTLCtermNF_2}{\mSTLCtype} (read similarly),
respectively.  We assume that
\(\STLCJudgeOne{\mSTLCterm_i}{\mSTLCtype}\) and
\(\STLCJudgeOne{\mSTLCtermNF_i}{\mSTLCtype}\) for \(i = 1,2 \).

\begin{defi}[Logical Relations for \STLC]
  The relations \STLCLogicalRelation{\STLCContext}%
  {\mSTLCterm_1}{\mSTLCterm_2}{\mSTLCtype} and
  \STLCLogicalRelationNF{\STLCContext}%
  {\mSTLCtermNF_1}{\mSTLCtermNF_2}{\mSTLCtype} are the least relation
  closed under the following rules:

 \STLCLogicalRelationRules
\end{defi}
The rule (\RSTLCLRKT) corresponds to (\AERKey) and means that the
number of keys to open a sealing with $\ell$ is at most one.  Although
we could give a more general definition of syntactic logical
relations, where the relation for type $\alpha_\ell$ is parameterized,
and prove the basic lemma for them below, but, in this paper, we do
not need such general settings and just take the restricted version
above for simplicity.
\begin{exa}
\label{ex:LESTLC}
\def\tempContext{%
\Dec{k}{\alpha_{\levL}}}
Take $\mSTLCterm_i$ such that \STLCJudge
{\tempContext}{\mSTLCterm_i}
{\alpha_{\levH} \to \bool} $(i = 1, 2)$.
They have normal forms by Strong Normalization.
Since there is no ``key'', that is, term of $\alpha_{\levH}$
under this variable context,
we cannot apply $\mSTLCterm_i$ to any terms of $\alpha_{\levH}$,
so \STLCLogicalRelation{\tempContext}{\mSTLCterm_1}{\mSTLCterm_2}
{\alpha_{\levH} \to \bool} by (\RSTLCLRTerm) and (\RSTLCLRFun).
This example almost corresponds to Example \ref{ex:LEone}.
In fact, we will translate $\GuardType{\levH}{\bool}$
and the observer level $\levH$, respectively, 
to $\alpha_{\levH} \to \bool$ and $\Dec{k}{\alpha_{\levH}}$,
in Section \ref{sec:Translation}.
\end{exa}

We write \(\delta\) for a simultaneous substitution of \STLC{} terms
for variables and
\STLCLogicalRelation{\STLCContext'}{\delta_1}{\delta_2}{\STLCContext} if
\(\dom(\delta_1) = \dom(\delta_2) = \dom(\STLCContext)\)
 and for any \(x \in \dom(\delta_1)\), 
\STLCLogicalRelation{\STLCContext'}{\delta_1(x)}{\delta_2(x)}{\STLCContext(x)}.
Then, the basic lemma is as follows:

\begin{lem}[Basic Lemma]
\label{lem:BasicLemma}
If $\STLCJudgeOne{\mSTLCterm}{\mSTLCtype}$ and
$\STLCLogicalRelation{\STLCContext'}%
   {\delta_1}{\delta_2}{\STLCContext}$,
then
$\STLCLogicalRelation{\STLCContext'}%
   {\delta_1(\mSTLCterm)}{\delta_2(\mSTLCterm)}{\mSTLCtype}$.
\end{lem}

For later use, we will prove a little generalized lemma as below, from
which the basic lemma above follows by reflexivity of $\equiv$ (Lemma
\ref{lem:axeq}).
\begin{lem}
 \label{lem:soundness}
 If $\AxiomaticEquiv{\STLCContext}%
      {\mSTLCterm_1}{\mSTLCterm_2}{\mSTLCtype}$
 and $\STLCLogicalRelation{\STLCContext'}%
       {\delta_1}{\delta_2}{\STLCContext}$,
 then $\STLCLogicalRelation{\STLCContext'}%
        {\delta_1(\mSTLCterm_1)}%
        {\delta_2(\mSTLCterm_2)}%
        {\mSTLCtype}$.
\end{lem}
\proof
By induction on the derivation of
\AxiomaticEquiv{\STLCContext}{\mSTLCterm_1}{\mSTLCterm_2}
{\mSTLCtype.} We show only the main cases.
Below, we write $\delta_1' \uplus \delta_2'$ for the union of two disjoint
substitutions $\delta_1'$ and $\delta_2'$ such that $\dom(\delta_1') \cap
\dom(\delta_2') = \emptyset$: $\dom(\delta_1' \uplus \delta_2') =
\dom(\delta_1') \cup \dom(\delta_2')$ and  $(\delta_1' \uplus \delta_2')(x) =
\delta_i' (x)$ if $x \in \dom(\delta_i')$.
\begin{mycase}[the last rule of the derivation is (\AERKey)]
Then, the last step of the derivation has a form
\[
 \infer
   {\AxiomaticEquiv{\STLCContext}%
     {\mSTLCterm_1}{\mSTLCterm_2}{\alpha_\ell}}
   {\STLCJudgeOne{\mSTLCterm_1}{\alpha_\ell}
   &
   \STLCJudgeOne{\mSTLCterm_2}{\alpha_\ell}}
\]
and $\mSTLCtype = \alpha_\ell$. By Substitution Property, Strong
 Normalization and Subject Reduction, there exists
$\mSTLCtermNF_i$ such that $\delta_i(M_i) \to^* \mSTLCtermNF_i$
and \STLCJudge{\STLCContext'}{\mSTLCtermNF_i}{\alpha_\ell}
$(i = 1, 2)$. So, since \STLCLogicalRelationNF{\STLCContext'}
{\mSTLCtermNF_1}{\mSTLCtermNF_2}{\alpha_\ell}
 by (\RSTLCLRKT), we get \STLCLogicalRelation{\STLCContext'}
 {\delta_1(\mSTLCterm_1)}{\delta_2(\mSTLCterm_2)}{\alpha_\ell}
 by (\RSTLCLRTerm).
\end{mycase}
\begin{mycase}[the last rule of the derivation is (\AERAbs)]
 Then, the last step of the derivation has a form
 \[
  \infer
    {\AxiomaticEquiv{\STLCContext}
      {\STLCLamExp{x}{\mSTLCtype_1}
                     {\mSTLCterm_1'}}
      {\STLCLamExp{x}{\mSTLCtype_1}
                     {\mSTLCterm_2'}}
      {\mSTLCtype_1 \to \mSTLCtype_2}}
    {\AxiomaticEquiv{\STLCContext,~\Dec{x}{\mSTLCtype_1}}
       {\mSTLCterm_1'}{\mSTLCterm_2'}{\mSTLCtype_2}}.
 \] 
and $\mSTLCterm_i = \STLCLamExp{x}{\mSTLCtype_1}{\mSTLCterm_i'}$
 $(i = 1, 2)$ and $\mSTLCtype = \mSTLCtype_1 \to \mSTLCtype_2$.
 By Strong Normalization, there exist $\mSTLCtermNF_i$
 such that $\delta_i(\mSTLCterm_i) \longrightarrow^* \mSTLCtermNF_i$
 $(i = 1, 2)$. Take arbitrary $\mSTLCterm_i''$ $(i = 1, 2)$ such that
 \STLCLogicalRelation{\STLCContext'}{\mSTLCterm_1''}{\mSTLCterm_2''}
 {\mSTLCtype_1}, then \STLCLogicalRelation{\STLCContext'}
 {\delta_1\uplus[\mSTLCterm_1''/x]}{\delta_2\uplus[\mSTLCterm_2''/x]}{\STLCContext
 \cup \{\Dec{x}{\mSTLCtype_1}\}.}
 By the induction hypothesis,
 \STLCLogicalRelation
 {\STLCContext'}{(\delta_1\uplus[\mSTLCterm_1''/x])(\mSTLCterm_1')}
 {(\delta_2\uplus[\mSTLCterm_2''/x])(\mSTLCterm_2')}{\mSTLCtype_2.}
 Since $\mSTLCtermNF_i~\mSTLCterm_i''$ have  the same normal
 forms as $(\delta_i\uplus[\mSTLCterm_i''/x])(\mSTLCterm_i')$
 for $i = 1, 2$, we have \STLCLogicalRelation{\STLCContext'}
 {\mSTLCtermNF_1~\mSTLCterm_1''}{\mSTLCtermNF_2~\mSTLCterm_2''}
 {\mSTLCtype_2}, and hence \STLCLogicalRelationNF{\STLCContext'}
 {\mSTLCtermNF_1}{\mSTLCtermNF_2}{\mSTLCtype_1 \to \mSTLCtype_2},
 so \STLCLogicalRelation{\STLCContext'}{\delta_1(\mSTLCterm_1)}
 {\delta_2(\mSTLCterm_2)}{\mSTLCtype_1 \to \mSTLCtype_2.}
\end{mycase}
\begin{mycase}[the last rule of the derivation is (\AERApp)]
 Then, the last step of the derivation has a form
 \[
  \infer
  {\AxiomaticEquiv{\STLCContext}%
     {\mSTLCterm_1'\,\mSTLCterm_1''}%
     {\mSTLCterm_2'\,\mSTLCterm_2''}%
     {\mSTLCtype_2}
  }
  {\AxiomaticEquiv{\STLCContext}%
     {\mSTLCterm_1'}%
     {\mSTLCterm_2'}%
     {\mSTLCtype_1\,\to\,\mSTLCtype_2}
   &
   \AxiomaticEquiv{\STLCContext}%
     {\mSTLCterm_1''}%
     {\mSTLCterm_2''}%
     {\mSTLCtype_1}
  }
 \]
 By the induction hypotheses, \STLCLogicalRelation{\STLCContext'}
 {\delta_1(\mSTLCterm_1')}{\delta_2(\mSTLCterm_2')}{\mSTLCtype_1 \to
 \mSTLCtype_2} and \STLCLogicalRelation{\STLCContext'}
 {\delta_1(\mSTLCterm_1'')}{\delta_2(\mSTLCterm_2'')}{\mSTLCtype_1.}
 By definition, there exist $\mSTLCtermNF_i$ 
 such that $\delta_i(\mSTLCterm_i') \longrightarrow^* \mSTLCtermNF_i$
 $(i = 1, 2)$ and \STLCLogicalRelationNF{\STLCContext'}{\mSTLCtermNF_1}
 {\mSTLCtermNF_2}{\mSTLCtype_1 \to \mSTLCtype_2}, and hence
 \STLCLogicalRelation{\STLCContext'}{\mSTLCtermNF_1~
 \delta_1(\mSTLCterm_1'')}{\mSTLCtermNF_2~\delta_2(\mSTLCterm_2'')}
 {\mSTLCtype_2.} Since $\delta_i(\mSTLCterm_i'~\mSTLCterm_i'')$
 have the same normal forms as $\mSTLCtermNF_i~
 \delta_i(\mSTLCterm_i'')$ for $i = 1, 2$, we have \STLCLogicalRelation
 {\STLCContext'}{\delta_1(\mSTLCterm_1'~\mSTLCterm_1'')}
 {\delta_2(\mSTLCterm_2'~\mSTLCterm_2'')}{\mSTLCtype_2.}
\qed
\end{mycase}

\begin{rem}
\label{rmk:STLCrefl}
Although the above logical relations for $\STLC$ are not reflexive in
general (for example $\Dec{x}{\mSTLCtype + \mSTLCtype} \not\seqsym x
\approx x : \mSTLCtype + \mSTLCtype$), 
we have $\STLCLogicalRelation{\STLCContext}{\mSTLCterm}
{\mSTLCterm}{\mSTLCtype}$ if all the types in $\STLCContext$
are of forms \(\mSTLCtype_1 \to \mSTLCtype_2 \to \dots \to \mSTLCtype_n \to
 \alpha_\ell\).  This is derived from Lemma
\ref{lem:BasicLemma} and the fact that
$\STLCLogicalRelation{\STLCContext}{x}{x}{\STLCContext(x)}$ 
if $\STLCContext(x) = \mSTLCtype_1 \to \mSTLCtype_2 \to \dots \to \mSTLCtype_n \to
 \alpha_\ell$, which can be proved by induction on $n$.
\end{rem}
%


\section{Translation}
\label{sec:Translation}
In this section, we define a formal translation from \dc{} to
$\STLC$ and its inverse.  Both translations are shown to preserve typing.  
\subsection{From \texorpdfstring{\dc{}}{dc} to \texorpdfstring{\STLC{}}{STLC}}
One of the main ideas of the translation, which closely follows Tse
and Zdancewic's translation from DCC to System
F~\cite{steve-translating-acm-conf,steve-translating-techreport}, is
to translate sealing of type \(\GuardType{\ell}{\mDMCtype}\) to a
function from the base type $\alpha_\ell$, which corresponds to
\(\ell\).  The sealed value can be extracted by passing a term of
$\alpha_\ell$ as an argument.  Intuitively, the term of $\alpha_\ell$
serves as a ``key'' for unsealing.

\begin{defi}[Translation of Types and Contexts]
$(\cdot)^\dag$ is a function from \dc{} types to 
\STLC{} types, defined by:
\TranslatingTypes
$(\cdot)^\dag$ is extended pointwise to contexts by:
 $\DMCContext^\dag
  = \{\Dec{x}{\mDMCtype^\dag} \,|\,
      \DMCDec{x}{\mDMCtype} \in \DMCContext\}$.
\end{defi}

Before describing the details of the translation, we give an example
for readers to grasp its intuition.
\begin{exa}
\label{ex:TrI}
We translate the \dc{} judgment
 \DMCJudge{\Dec{x}{\GuardType{\levL}{\bool}}}{\levH}{\DMCLappExp{x}{\levL}}{\bool}
to:
\[
 \STLCJudge{\Dec{x}{\alpha_\levL \to \bool},\,
            \Dec{c_{\levL\levL}}{\alpha_\levL \to \alpha_\levL},\,
            \Dec{c_{\levH\levH}}{\alpha_\levH \to \alpha_\levH},\,
            \Dec{c_{\levH\levL}}{\alpha_\levH \to \alpha_\levL},\,
            \Dec{k_\levH}{\alpha_\levH}}
  {x\,(c_{\levH\levL}\,k_\levH)}{\bool}.
\]
The first and last variable declarations are respectively translated
results of $\Dec{x}{\GuardType{\levL}{\bool}}$ and the observer level
$\levH$. The unsealing $\DMCLappExp{x}{\levL}$ is translated into the
application of $x$ to $c_{\levH\levL}\,k_\levH$ which corresponds to a
key for the unsealing, and where $c_{\levH\levL}$ coerces the key
$k_\levH$ for the observer level $\levH$ to that for $\levL$.  This
coercion is declared at the second last variable declaration.  The other
variables $c_{\levL\levL}$ and $c_{\levH\levH}$ are trivial coercions.
\end{exa}

Let $c$ be an injective partial map from pairs of levels to variables such
that $c_{\ell_2\,\ell_1}$ is defined if and only if $\ell_1 \sqsubseteq \ell_2$.
We take a finite mapping $\KC = \{
\Dec{c_{\ell_2\,\ell_1}}{\alpha_{\ell_2} \to \alpha_{\ell_1}} ~|~ \ell_1
\sqsubseteq \ell_2\}$ from variables to types, which corresponds to the
variable declarations \[
            \Dec{c_{\levL\levL}}{\alpha_\levL \to \alpha_\levL},\,
            \Dec{c_{\levH\levH}}{\alpha_\levH \to \alpha_\levH},\,
            \Dec{c_{\levH\levL}}{\alpha_\levH \to \alpha_\levL}
\] in Example \ref{ex:TrI}. Each variable
$c_{\ell_2\,\ell_1}$ represents a function to coerce a key for a higher
level to that for a lower. As like above, $\KC$ will be included in a
variable context for typing the translated terms.
Note that, if we let $\mathcal{L}$ be infinite, the domain of $\KC$
would be so, too, and hence we would have to extend the type judgments of
\STLC to allow an infinite context.  Such an extension would be easy
since only a finite number of variables can be used in a term.

The translation of \dc{} to \(\STLC\) is represented by
$\DMCtoSTLC{\DMCContext}{\concat{\sigma}{\barell}}
{\mDMCterm}{\mDMCtype}{\mSTLCterm}$, read ``\dc{} term $\mDMCterm$ of
type $\mDMCtype$ is translated to $\mSTLCterm$ under $\DMCContext$ and
$\sigma$,'' where $\sigma$ is an injective finite map from data levels
to variables.  In the example above, $\sigma$ is $\{\levH \mapsto
k_\levH\}$.  This mapping $\sigma$, whose domain represents the observer
level at which the \dc{} term is typed, records correspondence between
the data levels included in the observer level and variables that are used
as keys.  When typing the translated term in \STLC, those variables are
declared in the variable context (e.g., $\Dec{k_\levH}{\alpha_\levH}$ in
Example \ref{ex:TrI}), and hence, from usual conventions of \STLC, we
assume that the range of \(\sigma\) and the domains of \(\Gamma\) and
\(\KC\) are pairwise disjoint and that we can implicitly rename
variables in the range of $\sigma$, so that choices for key names do not
matter.

\begin{defi}[Translation of Terms]
The relation \DMCtoSTLC{\DMCContext}{\concat{\sigma}{\barell}}%
{\mDMCterm}{\mDMCtype}{\mSTLCterm} is defined as the least 
relation closed under the following rules:
\TranslationRules
Here, we write $\sigma\{\ell \mapsto k\}$ for a mapping from
$\dom(\sigma) \cup \{\ell\}$ to variables defined by: $\sigma\{\ell \mapsto
k\}(\ell) = k$; and $\sigma\{\ell \mapsto k\}(\ell') = \sigma(\ell')$ if \(\ell \neq
\ell'\). Note that $\ell$ may occur in the domain of $\sigma$.
\end{defi}
The translation of terms is easily derived from the translation rules
for types. In the last rule (\TransRLapp), a key for opening the sealing
is obtained from \(\sigma\) and a coercion---if
\(\DMCLappExp{\mDMCterm}{\ell}\) is well typed at the observer level
represented by \(\dom(\sigma)\), then \(\ell\) should be lower than
\(\dom(\sigma)\) and hence a coercion function should exist
 in $\KC$ to provide a key of $\ell$.
\begin{exa}
\label{ex:Tr}
Let $\levL$ and $\levH_1$ and $\levH_2$ be
data levels and suppose that $\levL$ is strictly lower
than both $\levH_1$ and $\levH_2$.
We can translate $\DMCJudge{\Dec{x}{\GuardType{\levL}{\bool}}}
{\levH_1, \levH_2}{\DMCLlamExp{\levH_1}{\DMCLappExp{x}{\levL}}}
{\GuardType{\levH_1}{\bool}}$ as follows:
\[
 \DMCtoSTLC{\Dec{x}{\GuardType{\levL}{\bool}}}
 {\{\levH_1 \mapsto k_1, \levH_2 \mapsto k_2\}}
 {\DMCLlamExp{\levH_1}{\DMCLappExp{x}{\levL}}}
 {\GuardType{\levH_1}{\bool}}
 {\STLCLamExp{k_1'}{\alpha_{\levH_1}\to\bool}
   {x\,K}}
\]
where $K$ is 
$c_{\levH_2\,\levL}\,k_2$ or
$c_{\levH_1\,\levL}\,k_1'$, but not $c_{\levH_1\,\levL}\,k_1$
because of the side condition of (\TransRLabs).
The resulting \STLC{} terms have type $\alpha_{\levH_1} \to \bool
(= \Dag{\GuardType{\levH_1}{\bool}})$ under context
\[
\STLCContext_0 \defeq \Dec{x}{\alpha_\levL\to\bool},\, \KC,\, \Dec{k_1}{\levH_1},\,
\Dec{k_2}{\levH_2}\ .
\]
\end{exa}
Well typed \dc{} terms can be translated
to well typed \STLC{} terms as in the theorem below.  Here, we write
$\Dag{\sigma}$ for the context defined by:
$\{\Dec{\sigma(\ell)}{\alpha_\ell} \mid \ell \in \dom(\sigma)\}$.
\begin{thm}[Translation Preserves Typing]
\label{theo:PTI}
 If \(\DMCJudgeOne{\mDMCterm}{\mDMCtype}\) and $\dom(\sigma) = \barell$, 
 then there exists a $\STLC$ term $\mSTLCterm$
 such that
 \(\DMCtoSTLC{\DMCContext}{\concat{\sigma}{\barell}}%
  {\mDMCterm}{\mDMCtype}{\mSTLCterm}\), and that
 \(\STLCJudge{\DMCContext^\dag,\,\KC,\, \Dag{\sigma}}%
  {\mSTLCterm}{\mDMCtype^\dag}\).
\end{thm}
\proof
 By induction on the derivation of \DMCJudgeOne{\mDMCterm}{\mDMCtype}.
 We show only the main cases:
\begin{mycase}[the last rule of the derivation is (\RDMCLabs)]
Then, \(\mDMCterm = \DMCLlamExp{\ell}{\mDMCterm_0}\) and 
\(\mDMCtype = \GuardType{\ell}{\mDMCtype_0}\) for some \(\mDMCterm_0\) and
\(\mDMCtype_0\).
 Take a fresh variable $k$ such that \(\ran(\sigma\{\ell \mapsto k\}) \cap
 \dom(\DMCContext) = \emptyset \). By the induction hypothesis,
 there exists $\mSTLCterm_0$ such that
 \DMCtoSTLC{\DMCContext}{\sigma\{\ell \mapsto k\}}%
         {\mDMCterm_0}{\mDMCtype_0}{\mSTLCterm_0}
 and
 \STLCJudge{\Dag{\DMCContext}, \KC,
          \Dag{(\sigma\{\ell \mapsto k\})}}%
         {\mSTLCterm_0}{\Dag{\mDMCtype_0}.}
 Note that $\Dag{(\sigma\{\ell \mapsto k\})} = \Dag{\sigma} \backslash
 \{\Dec{\sigma(\ell)}{\alpha_\ell}\} \cup \{\Dec{k}{\alpha_\ell}\}$.
 Hence, 
 \DMCtoSTLC{\DMCContext}{\sigma}{\DMCLlamExp{\ell}{\mDMCterm_0}}%
 {\GuardType{\ell}{\mDMCtype_0}}{\STLCLamExp{k}{\alpha_\ell}{\mSTLCterm_0}}
 and
 \STLCJudge{\Dag{\Gamma}, \KC, \Dag{\sigma}}{\STLCLamExp{k}{\alpha_\ell}%
  {\mSTLCterm_0}}{\Dag{\mDMCtype_0}} by (\STLCRTAbs) and weakening.
 \end{mycase}
\begin{mycase}[the last rule of the derivation is (\RDMCLapp)]
Then, \(\mDMCterm = \DMCLappExp{\mDMCterm_0}{\ell}\) for some 
\(\mDMCterm_0\).
 By the induction hypothesis, 
 there exists $\mSTLCterm_0$ such that
 \DMCtoSTLC{\DMCContext}{\sigma}%
         {\mDMCterm_0}{\GuardType{\ell}{\mDMCtype}}{\mSTLCterm_0}
 and
 \STLCJudge{\Dag{\DMCContext}, \KC, 
          \Dag{\sigma} }{\mSTLCterm_0}{\alpha_\ell \to \Dag{\mDMCtype}.}
 Note that $\ell \sqsubseteq \ell' \in \barell = \dom(\sigma)$, so
 \DMCtoSTLC{\DMCContext}{\sigma}{\DMCLappExp{\mDMCterm_0}{\ell}}%
         {\mDMCtype}{\mSTLCterm_0~(c_{\ell'\,\ell}\, \sigma(\ell'))}
 and
 \STLCJudge{\Dag{\DMCContext}, \KC,
          \Dag{\sigma} }{\mSTLCterm\,(c_{\ell'\,\ell}\,\sigma(\ell'))}{\Dag{\mDMCtype}.}
\end{mycase}
 The other cases are similar.
\qed
Note that, as we have seen in Example~\ref{ex:Tr}, the translation
result might not be unique since there might be many keys to be coerced
to one for some observer level in applying (\TransRLapp).  In fact, if
we can translate an unsealing term with some key included in $\sigma$,
where another higher key exists, then, another translation is also
possible by using the latter key instead of the former one, which may be
removed from $\sigma$. This fact is generalized as follows.
\begin{lem}
\label{lem:erase}
Assume that \DMCtoSTLC{\DMCContext}{\sigma\{\ell_1 \mapsto
k_1\}}{\mDMCterm}{\mDMCtype}{\mSTLCterm} and that $\ell_1 \sqsubseteq
\ell_2 \in \dom(\sigma)$. Then, there exists $\mSTLCterm'$ such that
\DMCtoSTLC{\DMCContext}{\sigma}{\mDMCterm}{\mDMCtype}{\mSTLCterm'} and,
if
\STLCJudge{\DMCContext^\dag,\,\KC,\,\sigma^\dag}{\mSTLCterm_1}{\alpha_{\ell_1},}
then
\AxiomaticEquiv{\DMCContext^\dag,\,\KC,\,\sigma^\dag}{[\mSTLCterm_1/k_1]\mSTLCterm}
{\mSTLCterm'}{\mDMCtype^\dag}.
The sizes of the derivations of the translations are the same.
\end{lem}
\proof By induction on the size of the derivation of
\DMCtoSTLC{\DMCContext}{\sigma\{\ell_1 \mapsto
k_1\}}{\mDMCterm}{\mDMCtype}{\mSTLCterm.}  Note that every occurrence
of $k_1$ in $\mSTLCterm$ appears as $c_{\ell_1\,\ell}\,k_1$ for some
$\ell$, since $k_1$ is always introduced by (\TransRLapp). Because
$\sigma$ has the higher key of $\alpha_{\ell_2}$ than $k_1$, we can
replace all the $c_{\ell_1\,\ell}\,k_1$ and remove all the occurrences
of $k_1$.  The last equivalence follows from (\AERKey).\qed

\subsection{From \texorpdfstring{\STLC{}}{STLC} to \texorpdfstring{\dc}{dc}}

We define the inverse translation, represented by\ \(
\STLCtoDMC{\DMCContext}{\concat{\sigma}{\barell}}%
{\mSTLCterm}{\mDMCtype}{\mDMCterm}\).  It is read ``$\STLC$ term
$\mSTLCterm$ of type $\mDMCtype^\dag$ under $\DMCContext^\dag$ and $\KC$ and
$\Dag{\sigma}$ is translated back to a \dc{} term \mDMCterm.''

\begin{defi}[Inverse Translation]
The relation \STLCtoDMC{\DMCContext}{\concat{\sigma}{\barell}}%
  {\mSTLCterm}{\mDMCtype}{\mDMCterm} is defined as the least relation
closed by the following rules:
\InverseTranslationRules
\end{defi}
In the rule (\InverseRLabsTwo), since we equate keys for the same data level
by (\AERKey) and (\RSTLCLRKT), we can replace the key $\sigma(\ell)$ by another $k$.
Note that, even if
\STLCJudge{\DMCContext^\dag,\,\KC,\,\Dag{\sigma}}{\mSTLCterm}{\mDMCtype^\dag,}
the inverse translation of \(\mSTLCterm\) is \emph{not} always possible.
However, we can give a sufficient condition for the inverse translation
to exist and show that the inverse translation also preserves typing:

\begin{thm}[Inverse Translation Preserves Typing]
\label{theo:PTII}
If all the subderivations of $\DMCContext^\dag, \KC,
\Dag{\sigma}$ $\seqsym \mSTLCterm : \mDMCtype^\dag$
satisfy SUbformula Property,
then there exists a \dc{} term $\mDMCterm$ such that
\(\DMCJudge{\DMCContext}{\dom(\sigma)}{\mDMCterm}{\mDMCtype}\)
and
\(\STLCtoDMC{\DMCContext}{\concat{\sigma}{\barell}}%
  {\mSTLCterm}{\mDMCtype}{\mDMCterm}\).
\end{thm}
\proof
By induction on the size of the derivation of
\STLCJudge{\DMCContext^\dag,\,\KC, %
  \Dag{\sigma}}{\mSTLCterm}{\mDMCtype^\dag.}
We show only the main cases:
\begin{mycase}[the last rule of the derivation is (\STLCRTAbs)]
Then, the last step of the derivation has a form
 \[
 \infer%
  {\STLCJudge{\Dag{\DMCContext}, \KC, \Dag{\sigma}}%
    {\STLCLamExp{x}{\mSTLCtype_1}{\mSTLCterm_0}}%
    {\mSTLCtype_1 \to \mSTLCtype_2}}
  {\STLCJudge{\Dag{\DMCContext}, \KC, \Dag{\sigma}, \Dec{x}{\mSTLCtype_1}}%
        {\mSTLCterm_0}{\mSTLCtype_2}},%
 \]
and $\Dag{t} = \mSTLCtype_1 \to \mSTLCtype_2$ and
 $M = \STLCLamExp{x}{\mSTLCtype_1}{\mSTLCterm_0}$.
 We have three subcases:
\begin{mysubcase}[$\mDMCtype = \mDMCtype_1 \to \mDMCtype_2$]
 Then, $\Dag{t_i} = \mSTLCtype_i (i = 1, 2)$ and
 \STLCJudge{\Dag{\DMCContext}, \Dec{x}{\Dag{\mDMCtype_1}}, \KC, \Dag{\sigma}}%
        {\mSTLCterm_0}{\Dag{\mDMCtype_2}},  all the subderivations
  of which also satisfy Subformula Property. So, by the
 induction hypothesis, there exists $\mDMCterm_0$ such that
 \DMCJudge{\DMCContext, \Dec{x}{\mDMCtype_1}}%
          {\dom(\sigma)}{\mDMCterm_0}{\mDMCtype_2}
 and
$ \STLCtoDMC{\DMCContext, \Dec{x}{\mDMCtype_1}}{\sigma}%
          {\mSTLCterm_0}{\mDMCtype_2}{\mDMCterm_0}$.
 Hence,
 \DMCJudge{\DMCContext}%
          {\dom(\sigma)}{\DMClamExp{x}{\mDMCtype_1}{\mDMCterm_0}}%
          {\mDMCtype_1 \to \mDMCtype_2}
 and
\(\STLCtoDMC{\DMCContext}{\sigma}%
          {\STLCLamExp{x}{\mSTLCtype_1}{\mSTLCterm_0}}%
          {\mDMCtype_1 \to \mDMCtype_2}%
          {\DMClamExp{x}{\mDMCtype_1}{\mDMCterm_0}}\).
\end{mysubcase}
\begin{mysubcase}[$\mDMCtype = \GuardType{\ell}{\mDMCtype_0}$ and 
$\ell \not\in \dom(\sigma)$] Then, $\mSTLCtype_1 = \alpha_\ell$ and
 $\mSTLCtype_2 = \mDMCtype_0^\dag$ and $\Dag{(\sigma\{\ell \mapsto x\})} =
 \Dag{\sigma} \cup \{\Dec{x}{\alpha_\ell}\}$ and
 \STLCJudge{\Dag{\DMCContext}, \KC, \Dag{(\sigma\{\ell \mapsto x\})}}
 {\mSTLCterm_0}{\Dag{\mDMCtype_0}}, all the subderivations of which also
 satisfy Subformula Property. So, by the induction hypothesis, there
 exists $\mDMCterm_0$ such that \DMCJudge{\DMCContext}{\dom(\sigma\{\ell
 \mapsto x\})}{\mDMCterm_0}{\mDMCtype_0} and
 \(\STLCtoDMC{\DMCContext}{\sigma\{\ell \mapsto x\}}
 {\mSTLCterm_0}{\mDMCtype_0}{\mDMCterm_0}\).  Since $\ell \not\in
 \dom(\sigma)$ and $\dom(\sigma\{\ell \mapsto x\}) = \dom(\sigma) \cup
 \{\ell\}$, it follows that 
 \DMCJudge{\DMCContext}{\dom(\sigma)}{\DMCLlamExp{\ell}{\mDMCterm_0}}
 {\GuardType{\ell}{\mDMCtype_0}} by (\RDMCLabs)
and \(\STLCtoDMC{\DMCContext}{\sigma}
 {\STLCLamExp{x}{\alpha_\ell}{\mSTLCterm_0}}
 {\GuardType{\ell}{\mDMCtype_0}}{\DMCLlamExp{\ell}{\mDMCterm_0}}\)
by (\InverseRLabsOne).
\end{mysubcase}
\begin{mysubcase}[$\mDMCtype = \GuardType{\ell}{\mDMCtype_0}$ and 
$\ell \in \dom(\sigma)$]
 Then, $\mSTLCtype_1 = \alpha_\ell$ and $\mSTLCtype_2 = \mDMCtype_0^\dag$ and
 $\Dag{(\sigma\{\ell \mapsto x\})} = \Dag{\sigma} \backslash
 \{\Dec{\sigma(\ell)}{\alpha_\ell}\}\cup \{\Dec{x}{\alpha_\ell}\}$
 and \STLCJudge{\Dag{\DMCContext}, \KC, \Dag{\sigma}, \Dec{x}{\alpha_\ell}}%
 {\mSTLCterm_0}{\Dag{\mDMCtype_0}.} By Substitution Property for \STLC{},
 \STLCJudge{\Dag{\DMCContext}, \KC, \Dag{(\sigma\{\ell \mapsto x\})}}%
 {[x/\sigma(\ell)]\mSTLCterm_0}{\Dag{\mDMCtype_0}} without changing the size of the
 derivation, all the subderivations of which also satisfy 
 Subformula Property. So, by the induction hypothesis, there exists a
 $\mDMCterm_0$ such that
 \DMCJudge{\DMCContext}{\dom(\sigma\{\ell \mapsto x\})}%
 {\mDMCterm_0}{\mDMCtype_0}
 and
 \(\STLCtoDMC{\DMCContext}{\sigma\{\ell \mapsto x\}}%
         {[x/\sigma(\ell)]\mSTLCterm_0}{\mDMCtype_0}{\mDMCterm_0}\).
Since
$\dom(\sigma\{\ell \mapsto x\}) = \dom(\sigma) \cup
 \{\ell\}$ and $\ell \in \dom(\sigma)$, it follows that 
 \DMCJudge{\DMCContext}{\dom(\sigma)}{\DMCLlamExp{\ell}{\mDMCterm_0}}%
        {\GuardType{\ell}{\mDMCtype_0}}
by (\RDMCLabs)
 and
 \(\STLCtoDMC{\DMCContext}{\sigma}%
         {\STLCLamExp{x}{\alpha_\ell}{\mSTLCterm_0}}%
         {\GuardType{\ell}{\mDMCtype_0}}{\DMCLlamExp{\ell}{\mDMCterm_0}}\)
by (\InverseRLabsTwo).
\end{mysubcase}
\end{mycase}
\begin{mycase}[the last rule of the derivation is (\STLCRTApp)]
 \def\tempcontext{\Dag{\DMCContext}, \KC, \Dag{\sigma}}%
Then, the last step of the derivation has a form
 \[
 \infer%
  {\STLCJudge{\tempcontext}%
     {\mSTLCterm_1\, \mSTLCterm_2}{\mSTLCtype_2}}%
  {\STLCJudge{\tempcontext}{\mSTLCterm_1}{\mSTLCtype_1 \to \mSTLCtype_2}
   &
   \STLCJudge{\tempcontext}{\mSTLCterm_2}{\mSTLCtype_1}}
 \]
and $\Dag{\mDMCtype} = \mSTLCtype_2$ and $\mSTLCterm = \mSTLCterm_1\,
 \mSTLCterm_2$. By Subformula Property, $\mSTLCtype_1$ and
 $\mSTLCtype_1 \to \mSTLCtype_2$ appear in $\Dag{\DMCContext} \cup \KC \cup
 \Dag{\sigma} \cup \Dag{\mDMCtype}$, hence, we have two cases about
 $\mSTLCtype_1$: $\mSTLCtype_1 = \alpha_\ell$ or  $\mSTLCtype_1 =
 \Dag{\mDMCtype_0}$ for some $\mDMCtype_0$.
 \begin{mysubcase}[$\mSTLCtype_1 = \alpha_\ell$]
  Then, $\mSTLCtype_1 \to \mSTLCtype_2 =
  \Dag{(\GuardType{\ell}{\mDMCtype})}$, by the induction hypothesis, there
  exists $\mDMCterm$ such that \DMCJudge{\DMCContext}{\dom(\sigma)}
  {\mDMCterm}{\GuardType{\ell}{\mDMCtype}} and
  \(\STLCtoDMC{\DMCContext}{\sigma}
  {\mSTLCterm_1}{\GuardType{\ell}{\mDMCtype}}{\mDMCterm}\).  Note that $\ell
  \sqsubseteq \dom(\sigma)$ since
  $\STLCJudge{\tempcontext}{\mSTLCterm_2}{\alpha_\ell}$.  So, it follows
  that \DMCJudge{\DMCContext}{\dom(\sigma)}
  {\DMCLappExp{\mDMCterm}{\ell}}{\mDMCtype} and
  \(\STLCtoDMC{\DMCContext}{\sigma} {\mSTLCterm_1~\mSTLCterm_2}
  {\mDMCtype}{\DMCLappExp{\mDMCterm}{\ell}}\) by (\RDMCLapp) and
  (\InverseRLapp).
  \end{mysubcase}
 \begin{mysubcase}[$\mSTLCtype_1 = \Dag{\mDMCtype_0}$]
  Then, $\mSTLCtype_1 \to \mSTLCtype_2 = \Dag{(\mDMCtype_0
  \to \mDMCtype_1)}$. By the induction hypotheses, we can easily show
  the conclusion.
 \end{mysubcase}
\end{mycase}
 For the cases where the last rule of the derivation is an elimination
 of a product or sum type, the proof is similar to the case of
 application. The rest of the proof is easy.
\qed
\begin{rem}
  In the above theorem, Subformula Property gives a sufficient
  condition to exclude ``junk'' terms such as \( (\STLCLamExp{x}%
  {\alpha_\ell \to \alpha_\ell}{()})%
  (\STLCLamExp{k}{\alpha_\ell}{k}) \).  Since
  \(\STLCLamExp{k}{\alpha_\ell}{k}\) has type \(\alpha_\ell \to
  \alpha_\ell\), no rules of inverse translation can be applied and
  the inverse translation will fail.  Its derivation, however, does
  not satisfy Subformula Property, so this is not a counterexample
  for the theorem above.  (In fact, its normal form can be translated
  back to a \dc{} term.)
\end{rem}
\begin{exa}
\label{ex:ITr}
We use the same settings as Example \ref{ex:Tr}.
\[
 \STLCtoDMC{\Dec{x}{\GuardType{\levL}{\bool}}}
 {\{\levH_1 \mapsto k_1, \levH_2 \mapsto k_2\}}
 {\STLCLamExp{k_1'}{\alpha_{\levH_1}\to\bool}
   {x\,K}}
 {\GuardType{\levH_1}{\bool}}
 {\DMCLlamExp{\levH_1}{\DMCLappExp{x}{\levL}}}
\]
where $K$ can be any term of type $\alpha_\levL$ under
context $\STLCContext_0,\,\Dec{k_1'}{\alpha_{\levH_1}\to\bool}$,
e.g, $c_{\levH_2\,\levL}\,k_2$ or
$c_{\levH_1\,\levL}\,k_1'$ or $c_{\levH_1\,\levL}\,k_1$.
\end{exa}

\section{Proof of Noninterference via Preservation of Logical Relations}
\label{sec:Proof}
In this section, we give an indirect proof of the noninterference
theorem, which is obtained as an easy corollary of the theorem that the
translation is sound and complete, that is, the logical relation for
\dc{} is preserved and reflected by the translation to \STLC{}.  The properties we
would expect are
\begin{quote}\it
     If \(\DMCtoSTLC{\cdot}{\concat{\sigma}{\barell}}%
          {\mDMCterm_i}{\mDMCtype}{\mSTLCterm_i}\)
	for $i = 1, 2$
     and \(\DMCLogicalRelation{\dom(\sigma)}%
	  {\mDMCterm_1}{\mDMCterm_2}{\mDMCtype}\),
     then \(\STLCLogicalRelation{\KC, \Dag{\sigma}}%
            {\mSTLCterm_1}{\mSTLCterm_2}{\mDMCtype^\dag}\),
\end{quote}
and its converse
\begin{quote}\it
     If \(\DMCtoSTLC{\cdot}{\concat{\sigma}{\barell}}%
          {\mDMCterm_i}{\mDMCtype}{\mSTLCterm_i}\) for $i = 1, 2$ and 
        \STLCLogicalRelation{\KC, \Dag{\sigma}}%
          {\mSTLCterm_1}{\mSTLCterm_2}{\mDMCtype^\dag},
     then
       \DMCLogicalRelation{\dom(\sigma)}{\mDMCterm_1}{\mDMCterm_2}{\mDMCtype}.
\end{quote}
It is not very easy, however, to prove them directly because logical
relations are defined by induction on types whereas the translations are
not.  Thus, following Tse and
Zdancewic~\cite{steve-translating-jfp-draft,steve-translating-acm-conf,steve-translating-techreport}, we introduce another
logical relation (called \emph{logical correspondence})
\(\LogicalCorrespondence{\concat{\sigma}{\barell}}%
{\mDMCterm}{\mSTLCterm}{\mDMCtype}\) over terms of \dc{} and \(\STLC\),
then prove that it includes (the graphs of) the translations of both
directions (Theorems~\ref{theo:inclusion} and
\ref{theo:includesinverse}).  Then, after showing that the logical
correspondence is full (Corollary~\ref{cor:fullness}), we finally
prove preservation of logical relations by logical correspondence and
reduce the noninterference theorem to Basic Lemma
(Lemma \ref{lem:BasicLemma}).

\subsection{Logical Correspondence and Its Fullness}
\label{subsec:LCandFull}
\begin{defi}[Logical Correspondence]
\label{def:LC}
The relations
\(\LogicalCorrespondence{\concat{\sigma}{\barell}}%
  {\mDMCterm}{\mSTLCterm}{\mDMCtype}\)
and 
\(\LogicalCorrespondenceNF{\concat{\sigma}{\barell}}%
  {\mDMCtermNF}{\mSTLCtermNF}{\mDMCtype}\),
where we assume that
$\DMCJudge{\cdot}{\dom(\sigma)}{\mDMCterm}{\mDMCtype}$ and
$\DMCJudge{\cdot}{\dom(\sigma)}{\mDMCtermNF}{\mDMCtype}$ and
$\STLCJudge{\KC, \sigma^\dag}{\mSTLCterm}{\mDMCtype^\dag}$ and
$\STLCJudge{\KC, \sigma^\dag}{\mSTLCtermNF}{\mDMCtype^\dag}$, are defined as
the least relation closed under the following rules:

\LogicalCorrespondenceRules

\end{defi}

Intuitively, \(\LogicalCorrespondence{\concat{\sigma}{\barell}}
{\mDMCterm}{\mSTLCterm}{\mDMCtype}\) means that \(\mDMCterm\) and
\(\mSTLCterm\) exhibit the same behavior from the viewpoint of an
observer at $\dom(\sigma)$.  The rule (\LCRLabel) for
\GuardType{\ell}{\mDMCtype} expresses the fact that the existence of
well typed $\mSTLCterm$ of $\alpha_\ell$ under $\KC$ and $\Dag{\sigma}$ is equivalent
to the fact that the level $\ell$ is lower than $\dom(\sigma)$.  In other
words, if \(\ell\) is not lower than \(\dom(\sigma)\), the premise is vacuously
true, representing that the observer cannot distinguish anything.
\begin{exa}
 Take \dc{} term $\mDMCterm$ and \STLC{} term $\mSTLCterm$ such
 that \DMCJudge{\cdot}{\levL}{\mDMCterm}{\GuardType{\levH}{\bool}}
 and \STLCJudge{\KC,\,\Dec{k}{\alpha_\levL}}{\mSTLCterm}
 {\alpha_\levH \to \bool}. By (\LCRTerm) and (\LCRLabel),
 \LogicalCorrespondence{\{\levL \mapsto k\}}{\mDMCterm}
 {\mSTLCterm}{\GuardType{\levH}{\bool}} because there is no term
 of type $\alpha_\levH$ under $\KC,\, \Dec{k}{\alpha_\levL}$.
 Compare this example with Examples \ref{ex:LEone} and \ref{ex:LESTLC}.
\end{exa}

Theorem~\ref{theo:closure} below shows that the logical
correspondences are closed under the composition with the logical
relations in $\STLC$.

\begin{thm}
\label{theo:closure}
If
$\LogicalCorrespondence{\concat{\sigma}{\barell}}%
  {\mDMCterm}{\mSTLCterm_1}{\mDMCtype}$  and
$\STLCLogicalRelation{\KC, \Dag{\sigma}}%
  {\mSTLCterm_1}{\mSTLCterm_2}{\mDMCtype^\dag}$,
then $\LogicalCorrespondence{\concat{\sigma}{\barell}}%
       {\mDMCterm}{\mSTLCterm_2}{\mDMCtype}$.
\end{thm}
\proof
By induction on the structure of $\mDMCtype$.
We show only the main cases:
\begin{mycase}[$\mDMCtype = \mDMCtype_1 \to \mDMCtype_2$]
 By definition, there exist $\mDMCtermNF$ and $\mSTLCtermNF_i$
 such that $\mDMCterm \longrightarrow^* \mDMCtermNF$
 and $\mSTLCterm_i \longrightarrow^* \mSTLCtermNF_i$ $(i = 1, 2)$
 and \LogicalCorrespondenceNF{\sigma}{\mDMCtermNF}{\mSTLCtermNF_1}%
 {\mDMCtype_1 \to \mDMCtype_2} and \STLCLogicalRelationNF{\KC, \Dag{\sigma}}%
 {\mSTLCtermNF_1}{\mSTLCtermNF_2}%
 {\Dag{\mDMCtype_1} \to \Dag{\mDMCtype_2}.} Take arbitrary $\mDMCterm_0$ and
 $\mSTLCterm_0$ such that \LogicalCorrespondence{\sigma}{\mDMCterm_0}%
 {\mSTLCterm_0}{\mDMCtype_1.} By definition, \LogicalCorrespondence%
 {\sigma}{\mDMCtermNF\,\mDMCterm_0}{\mSTLCtermNF_1\,\mSTLCterm_0}%
 {\mDMCtype_2.} Also, by Lemma \ref{lem:BasicLemma} (with Remark~\ref%
 {rmk:STLCrefl}), \STLCLogicalRelation{\KC, \Dag{\sigma}}{\mSTLCterm_0}%
 {\mSTLCterm_0}{\Dag{\mDMCtype_1}}, so, by definition,
 \STLCLogicalRelation{\KC, \Dag{\sigma}}{\mSTLCtermNF_1\,\mSTLCterm_0}%
 {\mSTLCtermNF_2\,\mSTLCterm_0}{\Dag{\mDMCtype_2}.} Applying the induction
 hypothesis for $\mDMCtype_2$, we have \LogicalCorrespondence{\sigma}%
 {\mDMCtermNF\,\mDMCterm_0}{\mSTLCtermNF_2\,\mSTLCterm_0}{\Dag{\mDMCtype_2}}
 and hence \LogicalCorrespondenceNF{\sigma}{\mDMCtermNF}{\mSTLCtermNF_2}%
 {\mDMCtype_1 \to \mDMCtype_2,} so \LogicalCorrespondence{\sigma}%
 {\mDMCterm}{\mSTLCterm_2}{\mDMCtype_1 \to \mDMCtype_2.}
\end{mycase}
\begin{mycase}[$\mDMCtype = \GuardType{\ell}{\mDMCtype_1}$]
 By definition, there exist $\mDMCtermNF$ and $\mSTLCtermNF_i$
 such that $\mDMCterm \longrightarrow^* \DMCLlamExp{\ell}{\mDMCtermNF}$
 and $\mSTLCterm_i \longrightarrow^* \mSTLCtermNF_i$ $(i = 1, 2)$ and
 \LogicalCorrespondenceNF{\sigma}{\DMCLlamExp{\ell}{\mDMCtermNF}}%
 {\mSTLCtermNF_1}{\GuardType{\ell}{\mDMCtype_1}} and \STLCLogicalRelation%
 {\KC, \Dag{\sigma}}{\mSTLCtermNF_1}{\mSTLCtermNF_2}%
 {\alpha_\ell \to \Dag{\mDMCtype_1}.} Take arbitrary $\mSTLCterm_0$ such 
 that \STLCJudge{\KC, \Dag{\sigma}}{\mSTLCterm_0}{\alpha_\ell.} By definition,
 \LogicalCorrespondence{\sigma}{\mDMCtermNF}{\mSTLCtermNF_1~\mSTLCterm_0}%
 {\mDMCtype_1} and \STLCLogicalRelation{\KC, \Dag{\sigma}}{\mSTLCterm_0}%
 {\mSTLCterm_0}{\alpha_\ell,} so, \STLCLogicalRelation{\KC, \Dag{\sigma}}%
 {\mSTLCtermNF_1~\mSTLCterm_0}{\mSTLCtermNF_2~\mSTLCterm_0}%
 {\Dag{\mDMCtype_1}.} Applying the induction hypothesis for
 $\mDMCtype_1$, we have \LogicalCorrespondence{\sigma}{\mDMCtermNF}%
 {\mSTLCtermNF_2~\mSTLCterm_0}{\mDMCtype_1} and hence %
 \LogicalCorrespondenceNF{\sigma}{\DMCLlamExp{\ell}{\mDMCtermNF}}%
 {\mSTLCtermNF_2}{\GuardType{\ell}{\mDMCtype_1},} so, %
 \LogicalCorrespondence{\sigma}{\mDMCterm}{\mSTLCterm_2}%
 {\GuardType{\ell}{\mDMCtype_1}.}
\qed
\end{mycase}

The next theorem shows that
these logical correspondences include
the graphs of the translation to \STLC{}.
We write \(\LogicalCorrespondence{\concat{\sigma}{\barell}}
     {\gamma}{\delta}{\DMCContext}\) if
\(\dom(\gamma) = \dom(\delta) = \dom(\DMCContext)\) and
\(\LogicalCorrespondence{\concat{\sigma}{\barell}}%
  {\gamma(x)}{\delta(x)}{\Gamma(x)}\) for any
\(x \in \dom(\Gamma)\).

\begin{thm}[Inclusion of Translation]
\label{theo:inclusion}
If $\DMCJudge{\DMCContext}{\dom(\sigma)}{\mDMCterm}{\mDMCtype}$ 
  and $\DMCtoSTLC{\DMCContext}{\concat{\sigma}{\barell}}%
  {\mDMCterm}{\mDMCtype}{\mSTLCterm}$ and
   $\LogicalCorrespondence{\concat{\sigma}{\barell}}%
     {\gamma}{\delta}{\DMCContext}$, 
then $\LogicalCorrespondence{\concat{\sigma}{\barell}}%
       {\gamma(\mDMCterm)}{\delta(\mSTLCterm)}{\mDMCtype}$.
\end{thm}
\proof
 By induction on the size of the derivation of
 $\DMCtoSTLC{\DMCContext}{\concat{\sigma}{\barell}}%
   {\mDMCterm}{\mDMCtype}{\mSTLCterm}$.
 We show only the main cases:
\begin{mycase}[the last translation rule of the derivation is
 (\TransRAbs)]
Then, the last step of the derivation has a form
 \[
 \infer
   {\DMCtoSTLC{\DMCContext}%
      {\concat{\sigma}{\barell}}%
      {\DMClamExp{x}{\mDMCtype_1}{\mDMCterm_0}}%
      {\mDMCtype_1 \to \mDMCtype_2}%
      {\STLCLamExp{x}{\mDMCtype_1^\dag}{\mSTLCterm_0}}}
   {\DMCtoSTLC{\DMCContext, \DMCDec{x}{\mDMCtype_1}}%
      {\concat{\sigma}{\barell}}%
      {\mDMCterm_0}{\mDMCtype_2}{\mSTLCterm_0}}.
 \]
 Take arbitrary $\mDMCterm_1$ and $\mSTLCterm_1$ such that
 $\LogicalCorrespondence{\sigma}{\mDMCterm_1}{\mSTLCterm_1}%
 {\mDMCtype_1,}$ then, \LogicalCorrespondence{\sigma}%
 {\gamma\uplus[\mDMCterm_1/x]}{\delta\uplus[\mSTLCterm_1/x]}{\DMCContext%
 \cup \{\Dec{x}{\mDMCtype_1}\}.} By the induction hypothesis,
 \LogicalCorrespondence{\sigma}{(\gamma\uplus[\mDMCterm_1/x])%
 (\mDMCterm_0)}{(\delta\uplus[\mSTLCterm_1/x])(\mSTLCterm_0)}{\mDMCtype_2.}
 Since $\gamma(\DMClamExp{x}{\mDMCtype_1}{\mDMCterm_0})~\mDMCterm_1$
 and $\delta(\STLCLamExp{x}{\mDMCtype_1^\dag}{\mSTLCterm_0})%
 ~\mSTLCterm_1$ have the same normal forms as
  $(\gamma\uplus[\mDMCterm_1/x])(\mDMCterm_0)$ and $(\delta\uplus%
 [\mSTLCterm_1/x])(\mSTLCterm_0)$, respectively, we have
 \LogicalCorrespondence{\sigma}%
 {\gamma(\DMClamExp{x}{\mDMCtype_1}{\mDMCterm_0})~\mDMCterm_1}%
 {\delta(\STLCLamExp{x}{\mDMCtype_1^\dag}{\mSTLCterm_0})~\mSTLCterm_1}%
 {\mDMCtype_2,} and hence \LogicalCorrespondence{\sigma}%
 {\gamma(\DMClamExp{x}{\mDMCtype_1}{\mDMCterm_0})}%
 {\delta(\STLCLamExp{x}{\mDMCtype_1^\dag}{\mSTLCterm_0})}%
 {\mDMCtype_1 \to \mDMCtype_2.}
\end{mycase}
\begin{mycase}[the last translation rule of the derivation is
 (\TransRApp)]
Then, the last step of the derivation has a form
 \[
  \infer%
    {\DMCtoSTLC{\DMCContext}{\concat{\sigma}{\barell}}%
       {\mDMCterm_1 \, \mDMCterm_2}%
       {\mDMCtype_2}%
       {\mSTLCterm_1 \, \mSTLCterm_2}
    }
    {\DMCtoSTLC{\DMCContext}{\concat{\sigma}{\barell}}%
       {\mDMCterm_1}{\mDMCtype_1 \to \mDMCtype_2}{\mSTLCterm_1}
     &
     \DMCtoSTLC{\DMCContext}{\concat{\sigma}{\barell}}%
       {\mDMCterm_2}{\mDMCtype_1}{\mSTLCterm_2}
    }.
 \]
 By the induction hypotheses, \LogicalCorrespondence{\sigma}%
 {\gamma(\mDMCterm_1)}{\delta(\mSTLCterm_1)}{\mDMCtype_1 \to
 \mDMCtype_2} and \LogicalCorrespondence{\sigma}%
 {\gamma(\mDMCterm_2)}{\delta(\mSTLCterm_2)}{\mDMCtype_1.}
 By Strong Normalization, $\gamma(\mDMCterm_1)$ and
 $\delta(\mSTLCterm_1)$ respectively have the unique normal forms
 $\mDMCtermNF$ and $\mSTLCtermNF$ such that \LogicalCorrespondenceNF%
 {\sigma}{\mDMCtermNF}{\mSTLCtermNF}{\mDMCtype_1 \to
 \mDMCtype_2.} By definition, we have \LogicalCorrespondence%
 {\sigma}{\mDMCtermNF~\gamma(\mDMCterm_2)}%
 {\mSTLCtermNF~\delta(\mSTLCterm_2)}%
 {\mDMCtype_2} and hence \LogicalCorrespondence{\sigma}%
 {\gamma(\mDMCterm_1~\mDMCterm_2)}{\delta(\mSTLCterm_1~\mSTLCterm_2)}{\mDMCtype_2.}
\end{mycase}
\begin{mycase}[the last translation rule of the derivation is
 (\TransRLabs)]
Then, the last step of the derivation has a form
 \[
  \infer%
    {\DMCtoSTLC{\DMCContext}{\concat{\sigma}{\barell}}%
       {\DMCLlamExp{\ell}{\mDMCterm_0}}%
       {\GuardType{\ell}{\mDMCtype_0}}%
       {\STLCLamExp{k}{\alpha_\ell}{\mSTLCterm_0}}%
    }%
    {\DMCtoSTLC{\DMCContext}%
       {\concat{\sigma\{\ell \mapsto k\}}%
          {\barell \cup \{\ell\}}
       }%
       {\mDMCterm_0}{\mDMCtype_0}{\mSTLCterm_0}
     \andalso
     \mbox{\(k\) fresh}
    }.
 \]
 Then, there exist $\mDMCtermNF$ and $\mSTLCtermNF$ such that 
 $\gamma(\mDMCterm_0) \longrightarrow^* \mDMCtermNF$ and
 $\delta(\STLCLamExp{k}{\alpha_\ell}{\mSTLCterm_0}) \longrightarrow^*
 \mSTLCtermNF$.
 Take arbitrary $\mSTLCterm_1$ such that \STLCJudge{\KC, \Dag{\sigma}}%
 {\mSTLCterm_1}{\alpha_\ell.} Then there exists $\ell' \in \dom(\sigma)$
 such that $\ell \sqsubseteq \ell'$ and,
 by Lemma \ref{lem:erase}, there exists $\mSTLCterm_0'$ such that
 \DMCtoSTLC{\DMCContext}{\sigma}{\mDMCterm_0}
{\mDMCtype_0}{\mSTLCterm_0'} and \AxiomaticEquiv
{\DMCContext^\dag,\,\KC,\,\Dag{\sigma}}{\mSTLCterm_0'}{[\mSTLCterm_1/k]\mSTLCterm_0}
{\mDMCtype_0^\dag}. So, by the induction hypothesis,
 \LogicalCorrespondence{\sigma}{\gamma(\mDMCterm_0)}
{\delta(\mSTLCterm_0')}{\mDMCtype_0}. Also, by Lemma \ref{lem:soundness}, we have 
 \STLCLogicalRelation{\KC, \Dag{\sigma}}
{\delta(\mSTLCterm_0')}{\delta([\mSTLCterm_1/k]\mSTLCterm_0)}{\Dag{\mDMCtype_0}}.
Since $\delta([\mSTLCterm_1/k]\mSTLCterm_0)$ and
 $\delta(\STLCLamExp{k}{\alpha_\ell}{\mSTLCterm_0}) \mSTLCterm_1$
have the same normal form,  \STLCLogicalRelation{\KC, \Dag{\sigma}}
{\delta(\mSTLCterm_0')}{\delta(\STLCLamExp{k}{\alpha_\ell}{\mSTLCterm_0})
 \mSTLCterm_1}{\Dag{\mDMCtype_0}}, and, applying Theorem
  \ref{theo:closure}, we get 
 \LogicalCorrespondence{\sigma}{\gamma(\mDMCterm_0)}%
 {\delta(\STLCLamExp{k}{\alpha_\ell}{\mSTLCterm_0})~\mSTLCterm_1}%
 {\mDMCtype_0,} hence \LogicalCorrespondence{\sigma}{\mDMCtermNF}%
 {\mSTLCtermNF~\mSTLCterm_1}{\mDMCtype_0,} so %
 \LogicalCorrespondenceNF{\sigma}{\DMCLlamExp{\ell}{\mDMCtermNF}}%
 {\mSTLCtermNF}{\GuardType{\ell}{\mDMCtype_0}.}
 Therefore \LogicalCorrespondence{\sigma}{\gamma(\DMCLlamExp{\ell}{\mDMCterm_0})}%
 {\delta(\STLCLamExp{k}{\alpha_\ell}{\mSTLCterm_0})}{\GuardType{\ell}{\mDMCtype_0}.}
\end{mycase}
\begin{mycase}[the last translation rule of the derivation is
 (\TransRLapp)]
  Assume that the last step of the derivation has a form
 \[
  \infer%
    {\DMCtoSTLC{\DMCContext}{\concat{\sigma}{\barell}}%
       {\DMCLappExp{\mDMCterm_1}{\ell}}%
       {\mDMCtype_1}
       {\mSTLCterm_1\,(c_{\ell'\,\ell}\,\sigma(\ell'))}
    }%
    {\DMCtoSTLC{\DMCContext}{\concat{\sigma}{\barell}}%
       {\mDMCterm_1}{\GuardType{\ell}{\mDMCtype_1}}{\mSTLCterm_1}
     & \ell' \in \dom(\sigma) & \ell \sqsubseteq \ell'
    }.
 \]
 By the induction hypothesis, \LogicalCorrespondence{\sigma}%
 {\gamma(\mDMCterm_1)}{\delta(\mSTLCterm_1)}{\GuardType{\ell}%
 {\mDMCtype_1}.} By definition, there exist $\mDMCtermNF$ and
 $\mSTLCtermNF$ such that $\gamma(\mDMCterm_1) \longrightarrow^*
 \DMCLlamExp{\ell}{\mDMCtermNF}$ and $\delta(\mSTLCterm_1)
 \longrightarrow^* \mSTLCtermNF$ and \LogicalCorrespondenceNF%
 {\sigma}{\DMCLlamExp{\ell}{\mDMCtermNF}}{\mSTLCtermNF}%
 {\GuardType{\ell}{\mDMCtype_1},} and hence \LogicalCorrespondence%
 {\sigma}{\mDMCtermNF}{\mSTLCtermNF~(c_{\ell'\,\ell}\,\sigma(\ell'))}{\GuardType{\ell}%
 {\mDMCtype_1}.} Since $\gamma(\DMCLappExp{\mDMCterm_1}{\ell})$
 and $\delta(\mSTLCterm_1~(c_{\ell'\,\ell}\,\sigma(\ell')))$
 respectively have the same normal forms as $\mDMCtermNF$ and
 $\mSTLCtermNF~(c_{\ell'\,\ell}\,\sigma(\ell'))$, we conclude \LogicalCorrespondence%
 {\sigma}{\gamma(\DMCLappExp{\mDMCterm_1}{\ell})}{\delta(\mSTLCterm_1
 ~(c_{\ell'\,\ell}\,\sigma(\ell')))}{\mDMCtype_1.}
\qed
\end{mycase}

It is slightly harder to show that the logical correspondence includes
the graphs of the inverse translation, since the inverse translation
is \emph{not} quite a (right) inverse of the translation to
\STLC: The inverse translation followed by the forward translation
may yield a term different from the original (see Examples \ref{ex:Tr} and \ref{ex:ITr}).
Fortunately, the difference is only slight: They differ only in
subterms of base types \(\alpha_\ell\) and are equivalent via
$\equiv$, thus logically related by Lemma \ref{lem:soundness}.
\begin{lem}
\label{lem:identity}
If $\STLCJudge{\DMCContext^\dag, \KC, \Dag{\sigma}}%
    {\mSTLCterm}{\mDMCtype^\dag}$ and
   $\STLCtoDMC{\DMCContext}{\concat{\sigma}{\barell}}%
     {\mSTLCterm}{\mDMCtype}{\mDMCterm}$ and
   $\DMCtoSTLC{\DMCContext}{\concat{\sigma}{\barell}}%
     {\mDMCterm}{\mDMCtype}{\mSTLCterm'}$,
then $\AxiomaticEquiv{\DMCContext^\dag, \KC, \Dag{\sigma}}%
       {\mSTLCterm}{\mSTLCterm'}{\mDMCtype^\dag}$.
\end{lem}
\proof
 By induction on the derivation of
 $\STLCtoDMC{\DMCContext}{\concat{\sigma}{\barell}}%
 {\mSTLCterm}{\mDMCtype}{\mDMCterm}$.
 We show only the main cases:
 \begin{mycase}[$\mDMCterm = \DMCLlamExp{\ell}{\mDMCterm_1}$
 and $\ell \not\in \dom(\sigma)$]
  Then, we can assume that the last steps of 
  the translation and the inverse respectively have
  the following forms:
\[
  \infer%
    {\STLCtoDMC{\DMCContext}{\concat{\sigma}{\barell}}%
       {\STLCLamExp{k}{\alpha_\ell}{\mSTLCterm_1}}%
       {\GuardType{\ell}{\mDMCtype_1}}%
       {\DMCLlamExp{\ell}{\mDMCterm_1}}%
    }%
    {\STLCtoDMC{\DMCContext}%
       {\concat{\sigma\{\ell \mapsto k\}}%
          {\barell \cup \{\ell\}}
       }%
       {\mSTLCterm_1}{\mDMCtype_1}{\mDMCterm_1}
    & \ell \not\in \dom(\sigma)
    }
\]
\[
   \infer%
     {\DMCtoSTLC{\DMCContext}{\concat{\sigma}{\barell}}%
        {\DMCLlamExp{\ell}{\mDMCterm_1}}%
        {\GuardType{\ell}{\mDMCtype_1}}%
        {\STLCLamExp{k_0}{\alpha_\ell}{\mSTLCterm_2}}%
     }%
     {\DMCtoSTLC{\DMCContext}%
        {\concat{\sigma\{\ell \mapsto k_0\}}%
           {\barell \cup \{\ell\}}
        }%
        {\mDMCterm_1}{\mDMCtype_1}{\mSTLCterm_2}
      &
      \mbox{\(k_0\) fresh}
     }. 
\]
 By renaming the bound variables, we can also take $k$
 as $k_0$. Hence, by the induction hypothesis, \AxiomaticEquiv%
 {\Dag{\DMCContext},\, \KC,\, \Dag{\sigma},\, \Dec{k}{\alpha_\ell}}{\mSTLCterm_1}%
 {\mSTLCterm_2}{\Dag{\mDMCtype_1},} so \AxiomaticEquiv%
 {\Dag{\DMCContext},\, \KC,\, \Dag{\sigma}}{\STLCLamExp{k}{\alpha_\ell}%
 {\mSTLCterm_1}}{\STLCLamExp{k}{\alpha_\ell}{\mSTLCterm_2}}%
 {\alpha_\ell \to \Dag{\mDMCtype_1}.}
 \end{mycase}
 \begin{mycase}[$\mDMCterm = \DMCLlamExp{\ell}{\mDMCterm_1}$
 and $\ell \in \dom(\sigma)$]
 Then, we can assume that the last steps of 
 the translation and the inverse respectively have
 the following forms:
\[
 \infer%
  {\STLCtoDMC{\DMCContext}{\concat{\sigma}{\barell}}%
     {\STLCLamExp{k}{\alpha_\ell}{\mSTLCterm_1}}%
     {\GuardType{\ell}{\mDMCtype_1}}%
     {\DMCLlamExp{\ell}{\mDMCterm_1}}%
  }%
  {\STLCtoDMC{\DMCContext}%
     {\concat{\sigma\{\ell \mapsto k\}}%
        {\barell}
     }%
     {[k/\sigma(\ell)]\mSTLCterm_1}{\mDMCtype_1}{\mDMCterm_1}
   &
   \ell \in \dom(\sigma)
  }
\]
\[
   \infer%
     {\DMCtoSTLC{\DMCContext}{\concat{\sigma}{\barell}}%
        {\DMCLlamExp{\ell}{\mDMCterm_1}}%
        {\GuardType{\ell}{\mDMCtype_1}}%
        {\STLCLamExp{k_0}{\alpha_\ell}{\mSTLCterm_2}}%
     }%
     {\DMCtoSTLC{\DMCContext}%
        {\concat{\sigma\{\ell \mapsto k_0\}}%
           {\barell \cup \{\ell\}}
        }%
        {\mDMCterm_1}{\mDMCtype_1}{\mSTLCterm_2}
      &
      \mbox{\(k_0\) fresh}
     }. 
\]
 By renaming the bound variables, we can also take $k$
 as $k_0$. Hence, by the induction hypothesis, \AxiomaticEquiv%
 {\Dag{\DMCContext},\, \KC,\, \Dag{\sigma} \backslash \{\Dec{\sigma(\ell)}{\alpha_\ell}\},~
  \Dec{k}{\alpha_\ell}}{[k/\sigma(\ell)]\mSTLCterm_1}{\mSTLCterm_2}%
 {\Dag{\mDMCtype_1}.} Since $k = k_0$ and $k_0$ is fresh, $k \neq
  \sigma(\ell)$, so, by weakening, \AxiomaticEquiv{\Dag{\DMCContext},\, \KC,\,
  ~\Dag{\sigma},~\Dec{k}{\alpha_\ell}}{[k/\sigma(\ell)]\mSTLCterm_1}%
 {\mSTLCterm_2}{\Dag{\mDMCtype_1}.} Applying Lemma \ref{lem:appxAE}
 and the transitivity of $\equiv$, we have \AxiomaticEquiv%
 {\Dag{\DMCContext},\,\KC,\,\Dag{\sigma},\,\Dec{k}{\alpha_\ell}}{\mSTLCterm_1}%
 {\mSTLCterm_2}{\Dag{\mDMCtype_1},} and hence \AxiomaticEquiv%
 {\Dag{\DMCContext},\,\KC,\,\Dag{\sigma}}{\STLCLamExp{k}{\alpha_\ell}%
 {\mSTLCterm_1}}{\STLCLamExp{k}{\alpha_\ell}{\mSTLCterm_2}}%
 {\alpha_\ell \to \Dag{\mDMCtype_1}.}
 \end{mycase}
 \begin{mycase}[$\mDMCterm = \DMCLappExp{\mDMCterm_1}{\ell}$]
 Then, we can assume that the last steps of 
 the translation and the inverse respectively have
 the following forms:
 \[
  \infer%
    {\STLCtoDMC{\DMCContext}{\concat{\sigma}{\barell}}%
       {\mSTLCterm_1\,\mSTLCterm_0}%
       {\mDMCtype_1}%
       {\DMCLappExp{\mDMCterm_1}{\ell}}
    }
    {\STLCtoDMC{\DMCContext}{\concat{\sigma}{\barell}}%
       {\mSTLCterm_1}{\GuardType{\ell}{\mDMCtype_1}}{\mDMCterm_1}
     \andalso
     \STLCJudge{\DMCContext^\dag,\,\KC,\, \Dag{\sigma}}%
       {\mSTLCterm_0}{\alpha_\ell}
    }
 \]
 \[
  \infer%
    {\DMCtoSTLC{\DMCContext}{\concat{\sigma}{\barell}}%
       {\DMCLappExp{\mDMCterm_1}{\ell}}%
       {\mDMCtype_1}
       {\mSTLCterm_2\,(c_{\ell'\,\ell}\,\sigma(\ell'))}
    }
    {\DMCtoSTLC{\DMCContext}{\concat{\sigma}{\barell}}%
       {\mDMCterm_1}{\GuardType{\ell}{\mDMCtype_1}}{\mSTLCterm_2}
     & \ell' \in \dom(\sigma) & \ell \sqsubseteq \ell'
    }.
 \]
 Hence, by the induction hypothesis, \AxiomaticEquiv{\Dag{\DMCContext},
 \, \KC,\, \Dag{\sigma}}{\mSTLCterm_1}{\mSTLCterm_2}{\alpha_\ell \to
 \Dag{\mDMCtype_1}.} Also, by definition, \AxiomaticEquiv{\Dag{\DMCContext},\,
 \KC,\, \Dag{\sigma}}{\mSTLCterm_0}{c_{\ell'\,\ell}\,\sigma(\ell')}{\alpha_\ell.} Hence
 \AxiomaticEquiv{\Dag{\DMCContext},\, \KC,\, \Dag{\sigma}}{\mSTLCterm_1
 ~\mSTLCterm_0}{\mSTLCterm_2~(c_{\ell'\,\ell}\,\sigma(\ell'))}{\Dag{\mDMCtype_1}.}
\qed
\end{mycase}

Then, we can show the following theorem:
\begin{thm}[Inclusion of Inverse Translation]
\label{theo:includesinverse}
   If $\STLCtoDMC{\DMCContext}{\concat{\sigma}{\barell}}%
        {\mSTLCterm}{\mDMCtype}{\mDMCterm}$ and
      $\LogicalCorrespondence{\concat{\sigma}{\barell}}%
        {\gamma}{\delta}{\DMCContext}$, 
  then $\LogicalCorrespondence{\concat{\sigma}{\barell}}%
         {\gamma(\mDMCterm)}{\delta(\mSTLCterm)}{\mDMCtype}$.
\end{thm}
\proof
By Theorem~\ref{theo:PTI}, there exists
$\mSTLCterm'$ such that
$\DMCtoSTLC{\DMCContext}{\concat{\sigma}{\barell}}%
  {\mDMCterm}{\mDMCtype}{\mSTLCterm'}$.
Then, by Lemma~\ref{lem:identity},
$\AxiomaticEquiv{\DMCContext^\dag, \KC, \Dag{\sigma}}%
  {\mSTLCterm}{\mSTLCterm'}{\mDMCtype^\dag}$.
Since 
$\STLCLogicalRelation{\KC, \Dag{\sigma}}%
  {\delta}{\delta}{\DMCContext^\dag}$
(using Remark~\ref{rmk:STLCrefl}),
$\STLCLogicalRelation{\KC, \Dag{\sigma}}%
  {\delta(\mSTLCterm)}%
  {\delta(\mSTLCterm')}%
  {\mDMCtype^\dag}$
by Lemma~\ref{lem:soundness}.
Then, by Theorem~\ref{theo:inclusion},
$\LogicalCorrespondence{\concat{\sigma}{\barell}}%
{\gamma(\mDMCterm)}{\delta(\mSTLCterm')}{\mDMCtype}$ and, by
Theorem~\ref{theo:closure} and the symmetricity of the logical
relation for $\STLC$, $\LogicalCorrespondence{\concat{\sigma}{\barell}}%
{\gamma(\mDMCterm)}{\delta(\mSTLCterm)}{\mDMCtype}$.
\qed

As a corollary, the logical correspondences is shown to be full.
\begin{cor}[Fullness of Logical Correspondences]
\label{cor:fullness}
If $\STLCJudge{\KC, \sigma^\dag}%
     {\mSTLCterm}{\mDMCtype^\dag}$,
then there exists a \dc{} term $\mDMCterm$ such that
$\LogicalCorrespondence{\sigma}%
  {\mDMCterm}{\mSTLCterm}{\mDMCtype}$.
\end{cor}
\proof 
By Theorem~\ref{theo:SubForm}, there exists $\mSTLCtermNF$ such
that $\mSTLCterm \longrightarrow^* \mSTLCtermNF$ and all the
subderivations of \STLCJudge{\KC,
\sigma^\dag}{\mSTLCtermNF}{\mDMCtype^\dag} satisfy Subformula
Property.  Applying Theorem~\ref{theo:PTII}, we get the inverse
$\mDMCterm$ of $\mSTLCtermNF$ such that
$\STLCtoDMC{\cdot}{\concat{\sigma}{\barell}}
{\mSTLCtermNF}{\mDMCtype}{\mDMCterm}$.  So, from
Theorem~\ref{theo:includesinverse}, $\LogicalCorrespondence{\sigma}
{\mDMCterm}{\mSTLCtermNF}{\mDMCtype}$, and hence
$\LogicalCorrespondence{\sigma} {\mDMCterm}{\mSTLCterm}{\mDMCtype}$.
\qed

\subsection{Preservation of Logical Relations}
\label{subsec:PLR}
By using the logical correspondence introduced above, we prove
that the logical relations are preserved by the logical correspondence.
\begin{thm}[Preservation of Equivalences]
\label{theo:PE} ~
 \begin{enumerate}
  \item\label{state:peq}
     If $\LogicalCorrespondence{\concat{\sigma}{\barell}}%
          {\mDMCterm_i}{\mSTLCterm_i}{\mDMCtype}$
	for $i = 1, 2$ and 
        $\DMCLogicalRelation{\dom(\sigma)}%
	  {\mDMCterm_1}{\mDMCterm_2}{\mDMCtype}$,
     then $\STLCLogicalRelation{\KC, \Dag{\sigma}}%
            {\mSTLCterm_1}{\mSTLCterm_2}{\mDMCtype^\dag}$.

  \item\label{state:pneq}
     Symmetrically,
     if $\LogicalCorrespondence{\concat{\sigma}{\barell}}%
          {\mDMCterm_i}{\mSTLCterm_i}{\mDMCtype}$
     for $i = 1, 2$ and 
     $\STLCLogicalRelation{\KC, \Dag{\sigma}}%
          {\mSTLCterm_1}{\mSTLCterm_2}{\mDMCtype^\dag}$,
     then
       $\DMCLogicalRelation{\dom(\sigma)}%
          {\mDMCterm_1}{\mDMCterm_2}{\mDMCtype}$.
 \end{enumerate}
\end{thm}
\proof
We prove both simultaneously by induction on the structure of $\mDMCtype$.
We show only the main cases:
\begin{mycase}[$\mDMCtype = \mDMCtype_1 \to \mDMCtype_2$]
To show (\ref{state:peq}), take arbitrary \(\mSTLCterm_1'\) and
\(\mSTLCterm_2'\) such that
$\STLCLogicalRelation{\KC, \Dag{\sigma}}%
   {\mSTLCterm_1'}{\mSTLCterm_2'}{\mDMCtype_1^\dag}$.
By fullness (Corollary~\ref{cor:fullness}),
there exist $\mDMCterm_i'$ such that
$\LogicalCorrespondence{\sigma}%
{\mDMCterm_i'}{\mSTLCterm_i'}{\mDMCtype_1}$ \((i=1,2)\),
and by the induction hypothesis (\ref{state:pneq})
for $\mDMCtype_1$, we have $\DMCLogicalRelation{\dom(\sigma)}%
{\mDMCterm_1'}{\mDMCterm_2'}{\mDMCtype_1}$.
Then, by definition,
there exist $\mDMCtermNF_i$ and $\mSTLCtermNF_i$
 such that
\(\mDMCterm_i \longrightarrow^* \mDMCtermNF_i\) and
\(\mSTLCterm_i \longrightarrow^* \mSTLCtermNF_i\) and
$\LogicalCorrespondence{\sigma}%
{\mDMCtermNF_i\,\mDMCterm_i'}%
{\mSTLCtermNF_i\,\mSTLCterm_i'}%
{\mDMCtype_2}$ for \(i=1, 2\),
and
$\DMCLogicalRelation{\dom(\sigma)}%
{\mDMCtermNF_1\,\mDMCterm_1'}%
{\mDMCtermNF_2\,\mDMCterm_2'}%
{\mDMCtype_2}$.
Applying the induction
hypothesis (\ref{state:peq}) for $\mDMCtype_2$ to them,
$\STLCLogicalRelation{\KC, \Dag{\sigma}}%
{\mSTLCtermNF_1\,\mSTLCterm_1'}%
{\mSTLCtermNF_2\,\mSTLCterm_2'}%
{\mDMCtype_2^\dag}$.
So we have
$\STLCLogicalRelationNF{\KC, \Dag{\sigma}}%
{\mSTLCtermNF_1}%
{\mSTLCtermNF_2}%
{\mDMCtype_1^\dag \to \mDMCtype_2^\dag}$, 
and hence
$\STLCLogicalRelation{\KC, \Dag{\sigma}}%
{\mSTLCterm_1}{\mSTLCterm_2}{\mDMCtype_1^\dag \to \mDMCtype_2^\dag}$.
The statement (\ref{state:pneq}) can be shown
 similarly, \emph{without} the fullness.
\end{mycase}

\begin{mycase}[$\mDMCtype = \GuardType{\ell}{\mDMCtype_1}$]
  To show (\ref{state:pneq}), we have two subcases: $\ell \sqsubseteq
  \dom(\sigma)$ or not.  If $\ell \sqsubseteq \ell' \in \dom(\sigma)$
  for some \(\ell'\), then, by definition, $\STLCLogicalRelation{\KC,
  \Dag{\sigma}}{c_{\ell'\,\ell}\,\sigma(\ell')}
  {c_{\ell'\,\ell}\,\sigma(\ell')}{\alpha_\ell}$.  Also, by definition,
  there exist $\mDMCtermNF_i$ and $\mSTLCtermNF_i$ such that
  $\mDMCterm_i \longrightarrow^* \DMCLlamExp{\ell} {\mDMCtermNF_i}$ and
  $\mSTLCterm_i \longrightarrow^* \mSTLCtermNF_i$ and
  $\LogicalCorrespondence{\sigma}{\mDMCtermNF_i}
  {\mSTLCtermNF_i\,(c_{\ell'\,\ell}\,\sigma(\ell'))}{\mDMCtype_1}$ for
  \(i=1, 2\), and $\STLCLogicalRelation{\KC,
  \Dag{\sigma}}{\mSTLCtermNF_1\,(c_{\ell'\,\ell}\,\sigma(\ell'))}
  {\mSTLCtermNF_2\,(c_{\ell'\,\ell}\,\sigma(\ell'))}{\mDMCtype_1^\dag}$.
  Applying the induction hypothesis (\ref{state:pneq}) for
  $\mDMCtype_1$, we have
  \DMCLogicalRelation{\dom(\sigma)}{\mDMCtermNF_1}{\mDMCtermNF_2}
  {\mDMCtype_1}, which is equivalent to \DMCLogicalRelationNF
  {\dom(\sigma)}{\mDMCtermNF_1}{\mDMCtermNF_2}{\mDMCtype_1}, so
  $\DMCLogicalRelation{\dom(\sigma)}{\mDMCterm_1}
  {\mDMCterm_2}{\GuardType{\ell}{\mDMCtype_1}}$. The case $\ell
  \not\sqsubseteq \dom(\sigma)$ is trivial. Showing (1) is easy since
  \STLCJudge{\KC, \Dag{\sigma}}{\mSTLCterm'}{\alpha_\ell} is equivalent
  to $\ell \sqsubseteq \dom(\sigma)$.  \qed
\end{mycase}

\subsection{Noninterference}
Then, we prove the noninterference theorem by reducing it to
Lemma~\ref{lem:BasicLemma}.
\begin{cor}[Noninterference]
\label{col:NI}
If $\DMCJudgeOne{\mDMCterm}{\mDMCtype}$ and
$\DMCLogicalRelation{\barell}{\gamma_1}{\gamma_2}{\DMCContext}$, then
$\DMCLogicalRelation{\barell}{\gamma_1(\mDMCterm)}{\gamma_2(\mDMCterm)}%
  {\mDMCtype}$.
\end{cor}
\proof
  Choose an arbitrary \(\sigma\) such that $\dom(\sigma) = \barell$ and
  \(\ran(\sigma) \cap \dom(\Gamma) = \emptyset\).  By Theorem
  \ref{theo:PTI}, $\DMCtoSTLC{\DMCContext}{\concat{\sigma}{\barell}}%
  {\mDMCterm}{\mDMCtype}{\mSTLCterm}$ and
  $\STLCJudge{\DMCContext^\dag, \KC, \Dag{\sigma}}%
  {\mSTLCterm}{\mDMCtype^\dag}$ for some \(M\).  Similarly, for any
  \(x\in \dom(\gamma_i)\) (\(i=1,2\)), there exists
  \(\mSTLCterm_{xi}\) such that
  $\DMCtoSTLC{\cdot}{\concat{\sigma}{\barell}}%
  {\gamma_i(x)}{\Gamma(x)}{\mSTLCterm_{xi}}$ and
  $\STLCJudge{\DMCContext^\dag, \KC, \Dag{\sigma}}%
  {\mSTLCterm_{xi}}{(\Gamma(x))^\dag}$.  Define \(\delta_i\) ($i=1,2$)
  as a simultaneous substitution such that \(\dom(\delta_i) = \dom(\gamma_i)\) and
  \(\delta_i(x) = \mSTLCterm_{xi}\) for \(x \in \dom(\delta_i)\).
  Then, by Theorem~\ref{theo:inclusion},
  $\LogicalCorrespondence{\concat{\sigma}{\barell}}%
  {\gamma_i}{\delta_i}{\Gamma}$ for $i = 1, 2$ and so
  $\LogicalCorrespondence{\concat{\sigma}{\barell}}%
  {\gamma_i(\mDMCterm)}{\delta_i(\mSTLCterm)}{\mDMCtype}$ for $i = 1,
  2$.
By applying Theorem~\ref{theo:PE}(1) to the assumption 
$\DMCLogicalRelation{\barell}{\gamma_1}{\gamma_2}{\DMCContext}$, we have
$\STLCLogicalRelation{\KC, \Dag{\sigma}}%
  {\delta_1}{\delta_2}{\DMCContext^\dag}$.  
Thus, by Lemma~\ref{lem:BasicLemma} (with Remark~\ref{rmk:STLCrefl}),
$\STLCLogicalRelation{\KC, \Dag{\sigma}}%
  {\delta_1(\mSTLCterm)}%
  {\delta_2(\mSTLCterm)}%
  {\mDMCtype^\dag}$.
Finally, by Theorem~\ref{theo:PE}(2),
$\DMCLogicalRelation{\barell}%
  {\gamma_1(\mDMCterm)}%
  {\gamma_2(\mDMCterm)}%
  {\mDMCtype}$.
\qed


\section{Comparison of DCC with \texorpdfstring{\dc}{dc}}
\label{sec:DCC}
In this section, we briefly review DCC~\cite{abadi99core} and discuss
why the translation from DCC to System F given by Tse and Zdancewic
\cite{steve-translating-acm-conf,steve-translating-techreport} is
neither full nor even sound.  Then, we discuss an extension \DCCpc{} of
DCC, which was proposed also by Tse and Zdancewic in order to make the
translation full
\cite{steve-translating-jfp-draft,steve-translating-acm-conf,steve-translating-techreport}.
Finally, we show that \DCCpc{} is equivalent to \dc{} by giving
translations between the two.

\subsection{DCC and Tse--Zdancewic's translation to System F}
DCC is an extension of the computational
\(\lambda\)-calculus~\cite{Moggi91IC} and uses monads indexed by
dependency levels (e.g., security levels, binding times) in order to
control the dependencies between computations.  The dependency levels
are partially ordered by \(\sqsubseteq\)\footnote{%
  In fact, the dependency levels were assumed be a
  lattice~\cite{abadi99core} but we do not need meets and joins in the
  following development.  } as in \dc; computation and data at a
higher level are permitted to depend on those at lower levels, but the
other direction of dependencies is forbidden.  Here, we briefly sketch
a simplified version of
DCC~\cite{steve-translating-acm-conf,steve-translating-techreport} (we
call it simply DCC), in which pointed types and recursion are omitted.

The syntax of DCC is defined as follows:
 \begin{align*}
 \mDPCtype &::=  \DPCunit \mid \mDPCtype\to\mDPCtype \mid
 \mDPCtype \times \mDPCtype \mid
 \mDPCtype + \mDPCtype \mid \ProtectType{\ell}{\mDPCtype} \\
  \mDPCterm &::= x \mid ()
        \mid \DPClamExp{x}{\mDPCtype}{\mDPCterm}
        \mid \mDPCterm\, \mDPCterm
        \mid \DPCpair{\mDPCterm}{\mDPCterm}
        \mid \DPCpiOne{\mDPCterm} \mid \DPCpiTwo{\mDPCterm}
        \mid \DPCiotaOne{\mDPCterm} \mid \DPCiotaTwo{\mDPCterm} \\
    & \mid \DPCcasExp{\mDPCterm}{\mDPCterm}{\mDPCterm}
      \mid \DPCLlamExp{\ell}{\mDPCterm}
      \mid \DPCbindExp{x}{\mDPCterm}{\mDPCterm}
 \end{align*}
 Roughly speaking, a monadic type \(T_\ell\; t\), the monadic unit
 \(\eta_\ell\; e\), and the bind operation \(\bind{x}{e_1}{e_2}\)
 correspond to sealing types \([t]_\ell\), sealing terms \([e]_\ell\),
 and unsealing terms \(e^\ell\), respectively.  The typing rule for
 \(\eta_\ell\) is as follows:
\typicallabel{}
\infrule
  {\Gamma \seqsym e : t}
  {\Gamma \seqsym \DPCLlamExp{\ell}{e} : \ProtectType{\ell}{t}}
Note that a type judgment of DCC lacks an observer level; instead,
the notion of protected types is introduced to prevent information
leakage and plays a key role in the following typing rule for
\texttt{bind}:
\infrule{
  \Gamma \seqsym e_1 : T_\ell\; t_1 \andalso
  \Gamma, x:t_1 \seqsym e_2 : t_2 \andalso
  \DPCPJudge{\ell}{t_2}
}{
  \Gamma \seqsym \bind{x}{e_1}{e_2} : t_2
}
\DPCProtectionRulesTwoColumn
Here, judgment \DPCPJudge{\ell}{t} is read as ``\(t\) is protected at
\(\ell\)''. Intuitively, this judgment means that observers only at a
level equal to or higher than \(\ell\) can obtain some bits of
information from the value of \(t\).

So, this rule ensures that the value of the whole term cannot be
examined at unrelated levels.  However, \texttt{bind} is restrictive in
the sense that \(\eta_\ell\) must be placed within the scope of \(x\) to
make \(t_2\) protected.  For example, the term \(\lambda
y:T_\ell\;\bool.\bind{x}{y}{\eta_\ell\; x}\) is given type \((T_\ell\; \bool) \to
(T_\ell\; \bool)\) while \iffull the term \fi \(\lambda
y:T_\ell\;\bool.\eta_\ell\;(\bind{x}{y}{x})\) cannot.  We will see that this
restriction is a source of the failure of fullness of the translation by
Tse and Zdancewic.  The other typing rules are the same as
\STLC.

The reduction rule for \texttt{bind} is
\(\bind{x}{\eta_\ell\,e_1}{e_2} \longrightarrow [e_1/x]e_2\). The
other reduction rules and the logical relations are essentially the same
as \dc{} except for the change from $\GuardType{\ell}{t}$ to $T_\ell\, t$.
The logical relations are indexed by an observer level (that is, a
finite set of data levels) rather than a single data level as in Tse
and
Zdancewic~\cite{steve-translating-acm-conf,steve-translating-techreport,steve-translating-jfp-draft}.
Although our definition is a straightforward extension of theirs, this
seems more natural for \DCCpc{} below, for the domains of the
relations are terms that are well typed at a given observer level.

A main idea of the translation by Tse and Zdancewic, which we have
followed in this paper, is to translate monadic types \(T_\ell\; t\) into
function types \(\alpha_\ell \to t\).  (Otherwise, type translation is
the same as ours.) Term translation, the details for which we refer to
\cite{steve-translating-acm-conf,steve-translating-techreport}, is more involved than our
translation, due to the complexity of
\texttt{bind} and protected types---we will see how they are expressed
in terms of our unsealing in the next section.

\subsection{Failure of Fullness and Soundness}

Now we explain why their translation is neither full nor sound.

Consider the DCC type \(t = T_\ell ((T_\ell\;\bool) \to \bool)\).  Then, any DCC
terms of this type is equivalent to (sealed) constant functions
\(\eta_\ell(\lambda x:T_\ell\;\bool. c)\) where \(c\) is either
\texttt{true} or \texttt{false}.  Note, in particular, that the term
\(e = \eta_\ell(\lambda y:T_\ell\;\bool. \bind{x}{y}{x})\) is \emph{ill
  typed} due to the restriction of the typing rule of \texttt{bind}.
As a result, the two terms
\[
  e_1 = \lambda f.\bind{f'}{f}{\eta_\ell\; (f'\;(\eta_\ell\; \texttt{true}))}
\]
and
\[
  e_2 = \lambda f.\bind{f'}{f}{\eta_\ell\; (f'\;(\eta_\ell\; \texttt{false}))}
\]
are logically related at the type \((T_\ell((T_\ell\;\bool) \to \bool)) \to
(T_\ell\;\bool)\) and level \(\ell\) since all we can pass to these
functions are the constant functions above and we cannot pass
non-constant functions such as $e$.

In System F, however, the translations of $e_1$ and $e_2$ are
\emph{not} logically related at type \(\alpha_\ell \to ((\alpha_\ell \to
\bool) \to \bool)\), which corresponds to the DCC type \(t\) above!
This is because they can be distinguished by applying them to the term
\(M = \lambda k:\alpha_\ell.  \lambda f\colon \alpha_\ell\to\bool. f k\),
which would correspond to \(e\).  

In short, there is no well typed DCC term that corresponds to \(M\)
(failure of fullness) and, as a result, the equivalence of \(e_1\) and
\(e_2\) is not preserved through the translation (failure of soundness).

\subsection{Tse and Zdancewic's Extension of DCC}
Interestingly, Tse and Zdancewic also noticed the restriction of the
typing for \texttt{bind} in DCC and proposed an extension of DCC by
introducing the notion of protection contexts (as a set of data
levels) to type judgments.  The typing rules for $\eta_\ell$ and
\texttt{bind} are changed as follows:
\typicallabel{D-Bind1}
\DPCEta
\DPCBindOne
\DPCBindTwo
\DPCContraction
The rule (\RDPCBindOne) is essential and just corresponds to the rule
(\RDMCLapp) of \dc. The rule (\RDPCProtected) means that a term of a
type protected by $\ell$ can be used by a user which does not have
$\ell$.  This extension allows terms like \(\lambda
y:T_\ell\;\bool.\eta_\ell(\bind{x}{y}{x})\) and \(\eta_\ell(\lambda
y:T_\ell\;\bool. \bind{x}{y}{x})\) to be well typed.  The rest of the
typing rules are the same as \dc.  The definitions of the reduction
rules and the logical relations are the same as DCC.

In the next subsection, we will show the three rules (\RDPCBindOne),
(\RDPCBindTwo), and (\RDPCProtected) are in fact derived forms in the
sense that \DCCpc{} and \dc{} are equivalent.

\begin{rem}
 \label{rem:dccpc}
  \DCCpc{} was proposed
  \cite{steve-translating-acm-conf,steve-translating-techreport} and simplified later by Tse and
  Zdancewic \cite{steve-translating-jfp-draft}.  
  In this paper, we use
  the simplified version with the following changes:
 \begin{itemize}
  \item We split the single typing rule for \texttt{bind} into the two rules.
  \item We add the rule (\RDPCProtected) above for the subject reduction
        property, which does not really hold in the original formulation,
        due to the reduction of \texttt{bind}.
 \end{itemize}
\end{rem}

\subsection{Isomorphisms between \texorpdfstring{\dco{}}{dco} and 
\texorpdfstring{\DCCpc}{DCCpc}}
\label{sec:represent}
\renewcommand{\DMCJudge}[4]{{\ensuremath{#1\,;\,#2\,\seqsym_{\lambda^{[\,]}}\,#3\,:\,#4}}}
\renewcommand{\DPCJudge}[4]{{\ensuremath{%
  #1\,;\,#2\,\seqsym_{\mathrm{DCC}_\mathrm{pc}}\,#3\,:\,#4}}}

We show correspondence between \dco{} and \DCCpc{} by giving a
translation \((\cdot)^{\bullet}\) from \dco{} to \DCCpc{} and its
inverse \((\cdot)^\circ\) and showing that both preserve logical
equivalences.  The inverse translation is inspired by Tse and
Zdancewic's translation from DCC to System
F~\cite{steve-translating-acm-conf,steve-translating-techreport}: We
obtain the inverse translation by comparing theirs with our full complete
translation from \dc{} to \STLC.  In what follows, we add subscripts
``\dco{}'' and ``\DCCpc'' to distinguish typing judgments of the two
calculi.

At the type level, both translations are easy---they just exchange
\(\GuardType{\ell}{\cdot}\) and \(T_\ell\):
\[
 ({\GuardType{\ell}{\mDMCtype}})^{\bullet}  \stackrel{\mathrm{def}}{=}
 \ProtectType{\ell}{(\mDMCtype^{\bullet})} \qquad
 (\ProtectType{\ell}{\mDMCtype})^{\circ} \stackrel{\mathrm{def}}{=}
 \GuardType{\ell}{\mDMCtype^{\circ}}
\]
(For other type constructors, both translations are trivial.)  At the
term level, \((\cdot)^{\bullet}\) is obvious---sealing and unsealing
can be straightforwardly expressed by \(\eta_\ell\) and \texttt{bind},
respectively:
\begin{align*}
  ({\DMCLlamExp{\ell}{\mDMCterm}})^{\bullet} & \stackrel{\mathrm{def}}{=}
  \DPCLlamExp{\ell}{(\mDMCterm^{\bullet})} \\
 (\DMCLappExp{\mDMCterm}{\ell})^{\bullet} & \stackrel{\mathrm{def}}{=}
  \DPCbindExp{x}{\mDMCterm^{\bullet}}{x}.
\end{align*}

The translation \((\cdot)^\circ\) for terms is more involved.  A main
difficulty is in the \texttt{bind} operator.  At first one might think
\DPCbindExp{x}{\mDPCterm_1}{\mDPCterm_2} can be expressed by $(\lambda
x. \mDPCterm_2^{\circ}) ~(\mDPCterm_1^{\circ})^\ell$, but, if
\DPCJudge{\DPCContext}{\barell}
{\DPCbindExp{x}{\mDPCterm_1}{\mDPCterm_2}}{\mDPCtype_2} is derived by
(\RDPCBindTwo), where $\ell \not\sqsubseteq \barell$ and
$\DPCPJudge{\ell}{\mDPCtype_2}$, then
$\DMCLappExp{(\mDMCterm_1^{\circ})}{\ell}$ is typable only at $\barell \cup
\{\ell\}$, which is \emph{strictly} higher than \(\barell\); so is $(\lambda
x. \mDPCterm_2^{\circ})~(\mDPCterm_1^{\circ})^\ell$.  Thus, this naive
translation does not quite preserve typing.

This problem is solved by observing that $\mDPCtype_2$ is protected at
\(\ell\) (i.e., $\DPCPJudge{\ell}{\mDPCtype_2}$).  First, we can seal $(\lambda x.
\mDPCterm_2^{\circ})~(\mDPCterm_1^{\circ})^\ell$ and derive
\DMCJudge{\DPCContext^{\circ}}{\barell} {\DMCLlamExp{\ell}{(\lambda x.
    \mDPCterm_2^{\circ})~(\mDPCterm_1^{\circ})^\ell}}
{\GuardType{\ell}{\mDPCtype_2^{\circ}}}.  Here, this sealing with $\ell$ is
redundant since $\mDMCtype_2$ is already protected by $\ell$.  In
fact, we can always eliminate such a sealing by applying an
\emph{anti-protection combinator}, defined below, of type
\(\GuardType{\ell}{\mDPCtype_2} \to \mDPCtype_2\).
\begin{defi}[Anti-Protection Combinators]
The set of closed terms \DMCUnprotect{\ell}{\mDMCtype} indexed by
protected types is
inductively defined as follows:
 \DPCPUnprotections
\end{defi}
These combinators intuitively mean that, for any \dco{} term $\mDMCterm$
of type \(\mDPCtype^\circ\) such that $\DPCPJudge{\ell}{\mDPCtype}$, the
sealing term $\DMCLlamExp{\ell}{\mDMCterm}$ can be unsealed at any
observer level.  This intuition is justified by the following
proposition:
\begin{prop}
\label{prop:unp}
 The following properties hold:
 \begin{enumerate}
  \item If \DPCPJudge{\ell}{\mDMCtype} and $\ell \sqsubseteq \barell$, then
  \DMCLogicalRelation{\barell}{\DMCUnprotect{\ell}{\mDMCtype}}
   {\DMClamExp{x}{\GuardType{\ell}{\mDMCtype^\circ}}{\DMCLappExp{x}{\ell}}}
   {\GuardType{\ell}{\mDMCtype^\circ} \to \mDMCtype^\circ}.
 \item If \DPCPJudge{\ell}{\mDMCtype} and $\ell \not\sqsubseteq \barell$, then
   \DMCLogicalRelation{\barell}{\mDMCterm_1}{\mDMCterm_2}{\mDMCtype^\circ}
   for any \dco{} terms $\mDMCterm_i$ such that
   $\DMCJudge{\cdot}{\barell}{\mDMCterm_i}{\mDMCtype^\circ}$ $(i = 1,
   2)$.  In particular, under the same assumptions, it follows that
   \DMCLogicalRelation{\barell}
   {\DMCUnprotect{\ell}{\mDMCtype}}{f}{\GuardType{\ell}{\mDMCtype^\circ} \to
     \mDMCtype^\circ} for any function $f$ such that
   \DMCJudge{\cdot}{\barell} {f}{\GuardType{\ell}{\mDMCtype^\circ} \to
     \mDMCtype^\circ}.
 \end{enumerate}
\end{prop}
\proof
By induction of the derivation of \DPCPJudge{\ell}{\mDMCtype}.
\qed
The second clause means that no term of a protected type
illegally leak any information.  A corresponding property has been
proved for DCC~\cite{abadi99core}.

Now we return to defining \((\cdot)^\circ\). For the \texttt{bind}
operator, we have two cases. (Strictly speaking, \((\cdot)^\circ\) is
defined by induction on the type derivation as in Section
\ref{sec:Translation}.)  If the last typing rule is (\RDPCBindOne),
the definition is just
\[
 (\texttt{bind}~ x = \mDMCterm_1~ \texttt{in}~ \mDMCterm_2)^{\circ}
 \stackrel{\mathrm{def}}{=}
 (\lambda x.~ \mDMCterm_2^{\circ})~
  \DMCLappExp{(\mDMCterm_1^{\circ})}{\ell},
\]
where $\mDMCterm_1$ and $\mDMCterm_2$ have types
$T_\ell\, \mDMCtype_1$ and $\mDMCtype_2$, respectively.  If it is
(\RDPCBindTwo), we can assume \DPCPJudge{\ell}{\mDMCtype_2} and
\[
 (\texttt{bind}~ x = \mDMCterm_1~ \texttt{in}~ \mDMCterm_2)^{\circ}
 \stackrel{\mathrm{def}}{=}
 \DMCUnprotect{\ell}{\mDMCtype_2}~ \GuardType{\ell}%
 {(\lambda x.~ \mDMCterm_2^{\circ})~ \DMCLappExp{(\mDMCterm_1^{\circ})}{\ell}}.
\]
Another interesting case is when the last step of the type 
derivation is 
\DPCContraction
The situation is similar to the case for (\RDPCBindTwo): the \DCCpc{}
type \(t\) is already protected at \(\ell\) and so \(\ell\) in the context
of the premise is redundant.  So, we obtain
$\DMCUnprotect{\ell}{\mDPCtype}~\DMCLlamExp{\ell}{\mDPCterm^{\circ}}$,
in which \(\mDPCterm^{\circ}\) is the translation from
$\DPCJudge{\DPCContext}{\barell \cup \{\ell\}}{\mDPCterm}{\mDPCtype}$.
For the other typing rules, the translation is trivial.  
For example,
\[
 (\DPCLlamExp{\ell}{\mDPCterm})^{\circ} \stackrel{\mathrm{def}}{=}
 \DMCLlamExp{\ell}{\mDPCterm^{\circ}}.
\]

Clearly, both translations preserve typing.  The following theorem
ensures that the translations preserve the logical relations, showing
\DCCpc{} and \dco{} are equivalent.
\begin{thm}[Preservation of Equivalences]
  $\mDPCterm_1 \approx_{\barell} \mDPCterm_2 : \mDPCtype$ in \DCCpc{} iff
  $\mDPCterm_1^{\circ} \approx_{\barell} \mDPCterm_2^{\circ} :
  \mDPCtype^{\circ}$ in \dco.  Also, $\mDPCterm_1^{\bullet}
  \approx_{\barell} \mDPCterm_2^{\bullet} : \mDPCtype^{\bullet}$ in
  \DCCpc{} iff $\mDPCterm_1 \approx_{\barell} \mDPCterm_2 : \mDPCtype$ in
  \dco.
\end{thm}
\proof
We just give a sketch, which is along a similar line as the proof of
Theorem \ref{theo:PE}.  First, like Definition \ref{def:LC}, we define
logical correspondences
$\LogicalCorrespondence{\barell}{\mDMCterm}{\mDPCterm'}{\mDMCtype}$ over
terms of \dco{} and \DCCpc{} indexed by observer levels $\barell$
(instead of finite maps, since both \dco{} and \DCCpc{} use the common
poset of data levels).  Then we show the inclusion of $(\cdot)^\circ$
and $(\cdot)^\bullet$, that is,
$\LogicalCorrespondence{\barell}{\mDMCterm}
{\mDMCterm^\circ}{\mDMCtype}$ and
$\LogicalCorrespondence{\barell}{\mDPCterm^\bullet}
{\mDPCterm}{\mDMCtype}$ (cf. Theorem \ref{theo:inclusion} and
\ref{theo:includesinverse}).  We use Proposition \ref{prop:unp} to prove
the former.  Finally, we show the preservation of the equivalences
(cf. Theorem \ref{theo:PE}) and, combining the inclusion of the
translations, get the result.
\qed


\section{Conclusion}
\label{sec:Conclusion}
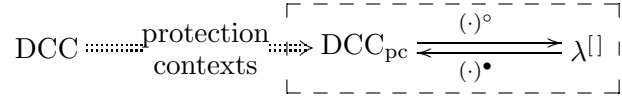
\begin{figure}
\[
\xymatrix@R-3ex{
&&&&
\save
[ll]+/r1.5pc/+/d0.4pc/.{[ddr]+/r1pc/+/u0.4pc/} *\frm{--}
\restore
&& \\
\mbox{DCC} \ar@{:>}[rrr]|{\txt{protection \\ contexts}} &&
&  \DCCpc  \ar@<0.4ex>[rr]^{(\cdot)^\circ}  &
 & \lambda^{[\,]}
\ar@<0.4ex>[ll]^{(\cdot)^\bullet} & \\
&&&&&&
}
\]
\caption{Relationship among DCC, \DCCpc, and \dc.}
\label{fig:four_calculi}  
\end{figure}

We have formalized noninterference for a typed \(\lambda\)-calculus
\dc{} by logical relations and proved it by
reducing it to the basic lemma of logical relation for $\STLC$ through
a translation of \dc{} to $\STLC$.  
Moreover, we have shown that \dc{} is equivalent to \DCCpc, an
extension of DCC with observer levels, as illustrated in
Figure~\ref{fig:four_calculi}: a dotted double arrow stands for a
language extension and the two systems (except DCC) in the dashed box
have sound and fully complete translations into \STLC.  In those
systems, dependency is captured by typability in \STLC{} through the
translations.

There have been presented many ways to prove noninterference theorems
for type-based dependency analyses for higher-order languages.  For
example, Heintze and Riecke~\cite{heintze-nevin98slam} and Abadi et
al.~\cite{abadi99core} showed the noninterference theorem for SLam by
using denotational semantics.  Pottier and Simonet
\cite{pottier-francois03info-flow-inference-ML} proved it for Core ML
with non-standard operational semantics.  Miyamoto and
Igarashi~\cite{miyaiga04modal}, in the study of a modal typed calculus
$\lambda_s^\Box$, showed that the noninterference theorem for certain
types can be easily proved only by using a simple nondeterministic
reduction system, although this system does not include recursion unlike
the others mentioned here.  

In comparison with these proofs, the proof technique presented in this
paper might seem overwhelming to show only noninterference.
Nevertheless, we believe it is still theoretically interesting since
the translation shows that the notion of dependency can be captured
only in terms of simple types and makes a comparison between
type-based dependency analyses easier.  

Practically, the translation might be a basis for implementing a
language with sealing by another language without it.  However, our
results rely on full reduction with commuting conversions, or strong
normalization, which cannot be assumed in real languages.  So, it
would be interesting future work to investigate how this proof
technique may be extended to richer languages with, for example,
recursion.  To add recursion, several difficulties have to be
overcome.  A first problem, as is already pointed out by Tse and
Zdancewic~\cite{steve-translating-jfp-draft,steve-translating-acm-conf,steve-translating-techreport},
is that a key of any data level can be ``forged'' by using recursion,
which allows a term of any type, and such forged keys enable any
observer to extract a sealed value illegally.  As suggested also by
Tse and Zdancewic, this problem may be solved by pointed types (or use
of Haskell's \texttt{seq}).  A second, more serious problem is that it
would be much harder to give an inverse translation: if the
translation is extended in a straightforward manner, then there will
be ``junk'' terms, such as some divergent terms not in the image of
the translation and, as a result, fullness would be lost.  We expect
some more significant work will be needed to solve these problems.


\section*{Acknowledgements}
Comments from anonymous referees helped up improve the final
presentation.  
We thank Masahito Hasegawa, Eijiro Sumii, Stephen Tse, and Steve
Zdancewic for discussions on this subject.  This work is supported in
part by Grant-in-Aid for Scientific Research (B) No.\ 17300003.

\bibliographystyle{plain}

\end{document}